\documentclass[aps, superscriptaddress,  twocolumn]{revtex4}
\usepackage{CJK}
\usepackage{mathtools}
\usepackage[normalem]{ulem}
\usepackage{graphicx}% Include figure files
\usepackage{dcolumn}% Align table columns on decimal point
\usepackage[mathlines]{lineno}
\usepackage{hyperref}
\usepackage{bm}% bold math
\usepackage{amssymb}
\usepackage{amsmath}
\usepackage{braket}
\usepackage{color}
\usepackage{xcolor}
\usepackage{soul} 
\usepackage{graphics}
\usepackage{bbm}
 \usepackage{xr}
\begin{document}
%\title{xxxx}
\title{Quantum Skyrmions in general quantum channels: topological noise rejection and the discretization of quantum information
}

\author{Robert de Mello Koch}
\affiliation{School of Science, Huzhou University, Huzhou 313000, China}
\affiliation{Mandelstam Institute for Theoretical Physics, School of Physics, University of the Witwatersrand, Private Bag 3, Wits 2050, South Africa}
\author{Bo-Qiang Lu}
\affiliation{School of Science, Huzhou University, Huzhou 313000, China}
\author{Pedro Ornelas}
\affiliation{School of Physics, University of the Witwatersrand, Private Bag 3, Wits 2050, South Africa}
\author{Isaac Nape}
\affiliation{School of Physics, University of the Witwatersrand, Private Bag 3, Wits 2050, South Africa}
\author{Andrew Forbes}
\affiliation{School of Physics, University of the Witwatersrand, Private Bag 3, Wits 2050, South Africa}
%
%
%%%%%%%%%%%%%%%%%%%%%%%%%%%%%%%%%%%%%%%%%%%%%%%%%%%
%%%% Abstract
%%%%%%%%%%%%%%%%%%%%%%%%%%%%%%%%%%%%%%%%%%%%%%%%%%%
\begin{abstract}
\noindent 
The topology of a pure state of two entangled photons is leveraged to provide a discretization of quantum information. Since discrete signals are inherently more resilient to the effects of perturbations, this discrete class of entanglement observables may offer an advantage against noise. Establishing this is the primary objective of this paper. We develop a noise model that exploits the specific form of such topological wave functions -- an entangled state of two photons, with one in an orbital angular momentum state and the other in a polarization state. We show that noise affecting both photons can be recast as a position-dependent perturbation affecting only the photon in the polarization state. This approach allows us to utilize both the language and concepts used in studying noisy qubits, as well as recent advances in quantum polarimetry. By adding noise to a finite-dimensional Hilbert space of polarization states, we can describe the noise using quantum operations expressed through appropriate Krauss operators, whose structure is determined by quantum polarimetry. For non-depolarizing noise, we provide an argument based on homotopic maps that demonstrates the topology's resilience to noise. For depolarizing noise, numerical studies using the quantum channel description show that the discrete entanglement signal remains completely resilient. Finally, we identify sources of local noise that can destabilize the topology. This foundational work establishes a framework for understanding how topology enhances the resilience of quantum information, directly impacting the distribution of information through entanglement in noisy environments, such as quantum computers and quantum networks.
\end{abstract}
\maketitle
\vspace{0.5cm}
\section{Introduction}

Entanglement is an intriguing aspect of quantum mechanics and has featured in numerous paradoxes such as those of Einstein-Podolsky-Rosen (EPR) \cite{EPR}, Hardy \cite{Hardy}, and Leggett \cite{Leggett2003} due to the nonlocal nature of quantum correlations. Presently, entangled states can be readily generated using various degrees of freedom of photons, e.g., polarisation \cite{kwiat1999ultrabright}, spatial/momentum \cite{mair2001entanglement, forbes2019quantum}, temporal/wavelength \cite{donohue2016spectrally, brendel1992experimental, brendel1999pulsed} and by mixing them to produce hybrid \cite{nagali2010generation, karimi2010spin, neves2009hybrid, nape2022all} and hyper \cite{ walborn2003hyperentanglement, gao2010experimental, xie2015harnessing, deng2017quantum} entangled states. This flexibility has made entanglement a key enabler for a wide range of quantum technologies, including, quantum key distribution \cite{durt2003security, yin2020entanglement, basso2021quantum}, quantum secret sharing \cite{
liao2014dynamic, hillery1999quantum}, quantum superdense coding  \cite{harrow2004superdense, barreiro2008beating} and quantum computing \cite{o2007optical, maring2024versatile}, promising enhanced secure communication networks and computing platforms.

However, entanglement is fragile and decays rapidly when subjected to spurious noise from the environment or imperfections from devices \cite{horodecki2003entanglement}. Known families of  channels that are responsible for this can be modelled as completely positive trace-preserving (CPTP) maps \cite{nielsen2001quantum}, that can include arbitrary combinations of unitary rotations, decoherence and dephasing channels,  resulting in actions such as a change of basis, phase errors or evolution from pure and entangled to noisy and mixed states \cite{kumar2003effect, lloyd1997capacity, liang2013quantum}. 

Various methods have been developed to counteract or mitigate these noise channels. The approaches either implement process tomography \cite{mohseni2008quantum} to find the channel and invert it \cite{valencia2020unscrambling}, or applies a distillation/purification  protocol that recovers the purity of the entangled states \cite{defienne2019quantum,huggins2021virtual, marshall2022distillation, bennett1996purification}. Alternatively, enhancing the robustness of entangled states against decay mechanisms by increasing the entanglement dimensionality \cite{ecker2019overcoming, almeida2007noise} or controlling the state symmetry \cite{azses2020identification} have also gained traction. However, these approaches come with several caveats; counteracting channel effects becomes increasingly difficult as the system size grows while enhancing robustness can increase noise tolerance but might not retain the same information capacity as an ideal system.

 Meanwhile, embedding topological features into hybrid entangled states has recently garnered attention \cite{ornelas2024non}, allowing for the preservation of topological invariants despite the presence of state decay. This has been achieved by imbuing the joint state of two entangled photons with an emergent topology that exists only as an entity of their joint state and is characterized by a topological wrapping number, i.e., the Skyrmion number ($N$). The integer $N$ remains unchanged under smooth deformations of the joint wave function. Consequently, this allows entangled states to be grouped into distinct topological classes, reminiscent of digitized signals, even though the wavefunctions of the same class may differ. Therefore, the entangled states remain in their topological classes regardless of the change in state. This notion has been shown for a special class of quantum channel, unitary amplitude-damping channels \cite{ornelas2024non}, but it remains unknown whether their topology is preserved in the presence of all known CPTP maps, in general. 

  Here, we address the full scope of the robustness question for Skyrmionic entangled states by identifying the family of CPTP maps (depolarizing and non-depolarizing) to which the topology of these states is robust and by unveiling the intricacies and nuances of dealing with maps to which the states are not robust. To this end, we introduce a generalized formalism, inspired by polarimetry, that can be applied to CPTP maps given the Krauss operators for the channels. For non-depolarizing sources of noise, our description employs a quantum channel with a single Kraus operator allowing for a rigorous argument based on homotopies between the noisy and noisefree density matrices therefore substantiating the topological noise rejection property. For decoherence channels where the operators are convex sums of Krauss operators, we show that although noise rejection is not simply established using homotopies, transitions between distinct topological numbers are not easily induced by noise, therefore enabling our scheme to benefit from a digitized topological encoding of the entangled states.

\section{The Skyrmion Wavefunction}\label{SWF}

The Skyrmion wave function is a pure state of two entangled photons. The first, photon $A$, occupies an orbital angular momentum state in a Hilbert space ${\cal H}_A$. The second, photon $B$, is in a polarization state in a Hilbert space ${\cal H}_B$. The two photon state belongs to the tensor product ${\cal H}_A\otimes{\cal H}_B$. A typical Skyrmion wave function takes the form
\begin{eqnarray}
|\psi\rangle&=&\int |\vec{r}_A\rangle\left(u_0^{l_1}(\vec{r}_A)|H\rangle_B+u_0^{l_2}(\vec{r}_A)e^{i\alpha}|V\rangle_B\right)\label{ourstate}
\end{eqnarray}
where $\alpha$ is an arbitrary phase, $l_1$ and $l_2$ are orbital angular momentum quantum numbers and the Laguerre-Gaussian modes are
\begin{eqnarray}
u^l_0(\rho,\phi)&=&\sqrt{2\over \pi |l|!}{1\over w_0}\left({\sqrt{2}\rho\over w_0}\right)^{|l|}e^{-{\rho^2\over w_0^2}}L_0^{|l|}\left({2\rho^2\over w_0^2}\right)e^{il\phi}\nonumber\end{eqnarray}
\noindent where $L$ is the generalised Laguerre polynomial and $w_0$ is the Gaussian width.  This state is described by the density matrix $\rho=|\psi\rangle\langle\psi|$. In terms of this density matrix, the Stokes parameters are
\begin{eqnarray}
  S_j(\vec{r}_A)&=&{\rm Tr}_B (\langle\vec{r}_A|\rho|\vec{r}_A\rangle\,\, \sigma_{Bj})
\qquad\qquad j=x,y,z \label{StokesP}
\end{eqnarray}
where $\sigma_{Bj}$ is a Pauli matrix $\sigma_j$ acting on ${\cal H}_B$
\begin{eqnarray}
\sigma_x=\left(\begin{array}{cc} 0 & 1 \\ 1 & 0 \\ \end{array}\right)\quad
\sigma_y=\left(\begin{array}{cc} 0 & -i \\ i & 0 \\  \end{array}\right)\quad
\sigma_z=\left(\begin{array}{cc} 1 & 0 \\ 0 & -1 \\ \end{array}\right)\label{PauliMats}
\end{eqnarray}
Normalize the Stokes parameters to define a unit vector ($\tilde{S}_i$) describing a point on a two sphere S$^2$ at each $\vec{r}_A$
\begin{eqnarray}
\tilde{S}_j=\alpha S_j\qquad j=x,y,z\qquad\qquad\sum_j \tilde{S}_j\tilde{S}_j=1
\label{NStokesP}
\end{eqnarray}
where $\alpha$ is fixed by the normalization condition $\vec{\tilde{S}}\cdot\vec{\tilde{S}}=1$. 

The map $\tilde{S}_j(x,y)$ plays a central role. It's base space is the set of all points on a two dimensional plane, which itself maps to a two sphere after compactifying and employing a stereographic map. Since $\tilde{S}_j(x,y)$ is a normalized vector, the target space of this map is also a two sphere - the Poincar\'e sphere. Consequently $\tilde{S}_j(x,y)$ maps a two sphere to a two sphere. Any map of this type has an associated winding number which counts how many times the first two sphere is wrapped around the second. In the current context the winding number is called the Skyrmion number and it may be computed as follows
\begin{eqnarray}
  N&=&{1\over 4\pi}\int_{-\infty}^\infty\int_{-\infty}^\infty\Sigma_z(x,y)\,dx\, dy\cr\cr
  \Sigma_z(x,y)&=&\epsilon_{pqr}\tilde{S}_p{\partial \tilde{S}_q\over\partial x}{\partial \tilde{S}_r\over\partial y}\label{SkNumb}
\end{eqnarray}
A product wave function, which is therefore not entangled, would map all points in the base space to a single point on the Poincar\'e sphere. This is a map with winding number zero. A non-zero winding number indicates that the map cannot be reduced to a trivial (constant) map without tearing or discontinuity. Thus, the map wraps around the target space in a non-trivial way, which cannot be undone by any smooth deformation. This non-trivial wrapping is simultaneously a manifestation of non-trivial topology and the fact that the wave function is not a product i.e. that it is entangled. This observation reveals a profound connection between topology and entanglement and it is the winding number $N$ that provides the discretization of entanglement. The key question we address in this study, is how noise affects the value of $N$.

For the wave function given above, we easily verify that
\begin{eqnarray}
N&=&l_1-l_2
\end{eqnarray}

\section{Noise Model}\label{NoiseModel}

\begin{figure*}[t!]
\includegraphics[width=1\linewidth]{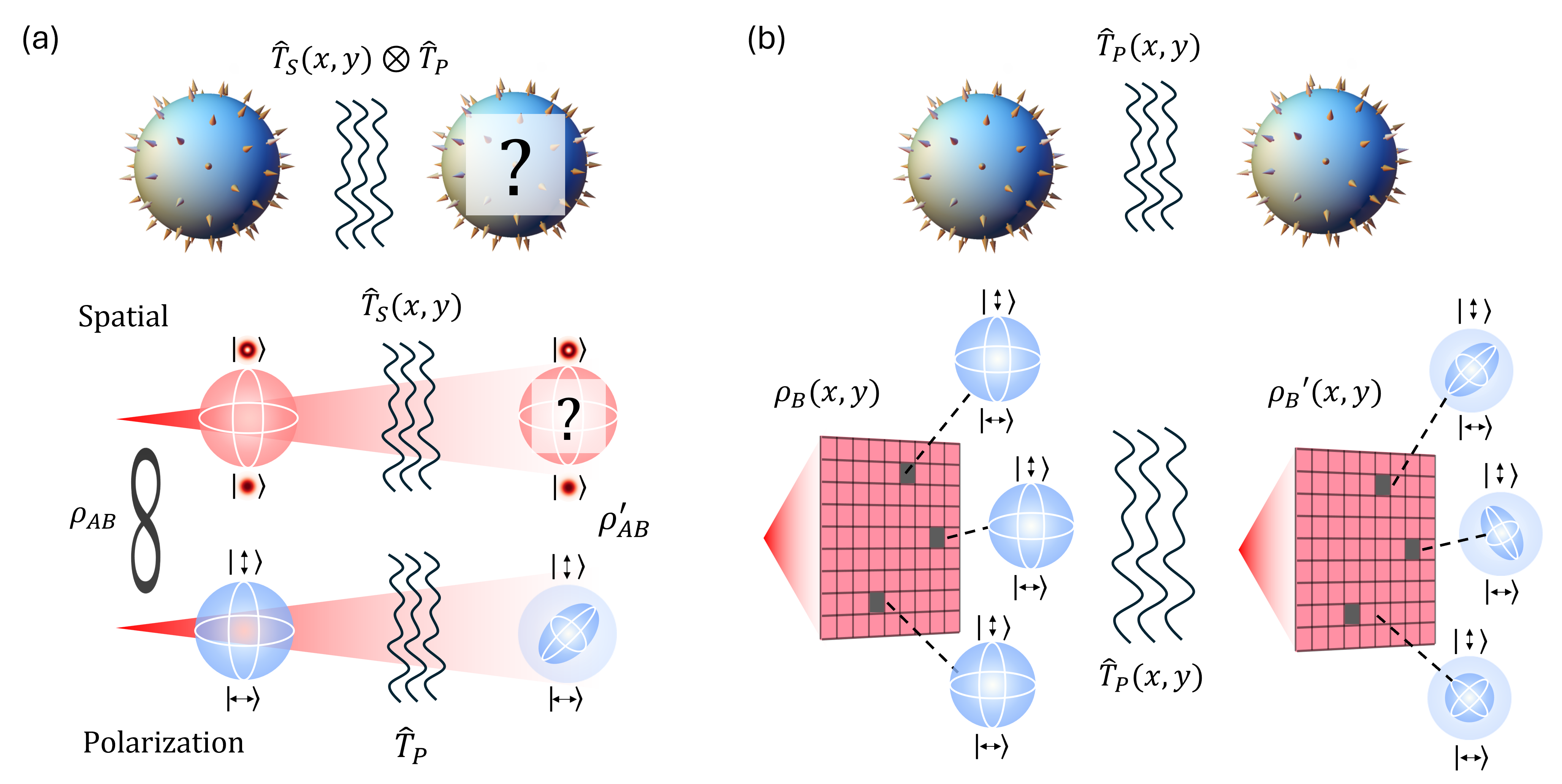}
\caption{\textbf{Skyrmion noise model.} Photons A and B are entangled in space and polarization, respectively, such that their entanglement exhibits non-trivial topological features. (a) Investigating the general behaviour of such features under the influence of noise requires passing photon A (spatial) through a noise channel ($\hat{T}_S(x,y)$) which modulates the amplitude and phase of the photon at different points in space and photon B (polarization) through a channel ($\hat{T}_P$) which transforms it's polarization. %Whilst arbitrary transformations on 2D polarization hilbert spaces are well understood, the analysis of transformations on infinite dimensional spatial hilbert spaces remain challenging. Instead (b) we consider an equivalent scenario whereby the spatially varying noise modulates polarization differently at different points in space.
(b) By exploiting the explicit form of the skyrmion wavefunction a more convenient model can be used which considers the noise as only influencing the polarization of photon B at different points in space ($\hat{T}_P(x,y)$), i.e., a position-dependent source of noise acting only on the polarization degree of freedom.}
\label{fig:concept}
\end{figure*}

In this section, we aim to introduce a noise model that is comprehensive enough to accurately reflect real-world physical systems. To begin, we might consider a scenario where photons A and B are analyzed separately, as depicted in Fig.~\ref{fig:concept} (a). This approach involves understanding how the states of photons A and B transform under various types of spatial and polarization noise, respectively. Given the complexity of the infinite-dimensional Hilbert space for photon A, this task can be daunting. Instead, by leveraging the explicit form of the Skyrmion wave function (\ref{ourstate}), we simplify our analysis. This approach confines the effect of noise to the polarization state of photon B, streamlining the problem. To ensure that our description remains comprehensive, we must allow for position-dependent noise, as illustrated in Fig.~\ref{fig:concept} (b). Our model captures the impact of noise on the density matrix, and accounting for its effect on the Skyrmion number is then a straightforward extension using formulas (\ref{StokesP}), (\ref{NStokesP}), and (\ref{SkNumb}).

Noise can, in general, affect the states of both photons in the entangled Skyrmion state (\ref{ourstate}). To account for this, consider noise induced perturbations of the form \footnote{We do not consider fluctuations of the relative phase, but these can easily be included. Our motivation for ignoring this possibility is simply the fact that the Skyrmion number is completely insensitive to $\alpha$.}
\begin{eqnarray}
u_0^{l_i}(\vec{r}_A)&\to&u_0^{l_i}(\vec{r}_A)+u_\delta^i(\vec{r}_A)\qquad i\,=\,1,2\cr\cr
|P\rangle_B&\to&|P\rangle_B+|\delta P\rangle_B\qquad\quad P\,=\, H,V
\label{gennoise}
\end{eqnarray}
These perturbations are not all independent since the trace of the density matrix must always be preserved. This gives a constraint written as an equation for the integral over the fluctuations. Since it is not very constraining or illuminating we don't explore it further. The most practical way to proceed is simply to ignore the constraint and then normalize the trace of the noisy density matrix to 1 before it is interpreted. Adding the noise perturbations we obtain the following noisy state
\begin{eqnarray}
|\psi\rangle&=& \int |\vec{r}_A\rangle\left(u_0^{l_1}(\vec{r}_A)(|H\rangle_B+|\Delta H\rangle_B)\right.\cr
&&\left.\qquad\qquad +\,\,u_0^{l_2}(\vec{r}_A)e^{i\alpha}(|V\rangle_B+|\Delta V\rangle_B)\right)\label{nicenoise}
\end{eqnarray}
where
\begin{eqnarray}
\Delta |H\rangle_B&=&{u_\delta^1(\vec{r}_A)\over u_0^{l_1}(\vec{r}_A)}|H\rangle_B+|\delta H\rangle_B+{u_\delta^1(\vec{r}_A)\over u_0^{l_1}(\vec{r}_A)}|\delta H\rangle_B\cr\cr
\Delta |V\rangle_B&=&{u_\delta^2(\vec{r}_A)\over u_0^{l_2}(\vec{r}_A)}|V\rangle_B+|\delta V\rangle_B+{u_\delta^2(\vec{r}_A)\over u_0^{l_2}(\vec{r}_A)}|\delta V\rangle_B\nonumber
\end{eqnarray}
We have manipulated the noisy state (\ref{nicenoise}) so that it looks as if photon A is completely unaffected by noise. The price for this is that the polarization vectors of photon B suffer position dependent noise. This manipulation relies heavily on the detailed form of the Skyrmion wave function (\ref{ourstate}). This is not a loss of generality, since the Skyrmion number itself is tied to this state. The clear advantage of this point of view is that we are now adding noise to the finite dimensional Hilbert space of polarization states of photon B. This immediately allows us to profitably employ both the language and ideas developed to study noisy qubits, as well as recent advances in quantum polarimetry. Specifically, the fact that we add noise to a finite dimensional Hilbert space allows us to phrase the discussion in terms of noise channels, using the notion of a quantum operation ${\cal E}$ which maps density matrices to density matrices. The operator sum representation of ${\cal E}$ is
\begin{eqnarray}
{\cal E}(\rho)&=&\sum_i E_i\rho E_i^\dagger\qquad\qquad\sum_i E_i^\dagger E_i={\bf 1}
\end{eqnarray}
where the $E_i$ are known as Krauss operators. Quantum polarimetry will determine the structure of the Krauss operators. 

\section{Quantum Theory of Polarimetry}\label{QPol}

In this section we review relevant aspects of the framework of polarimetry in terms of quantum mechanical operators acting on quantum states of light, as developed in \cite{goldberg2020quantum}. Polarization is characterized by four Stokes parameters, and polarimetry involves the $4\times 4$ Mueller matrix, which governs the linear transformations of these parameters under various optical elements. The approach in \cite{goldberg2020quantum} extends this to the quantum domain by describing quantum channels that correspond to Mueller matrices. For the quantum Stokes operators to align with the predictions of classical polarimetry, they must transform according to Mueller calculus, irrespective of the quantum state. A few basic quantum operations form the building blocks of quantum polarimetry. These building blocks are combined into composite quantum operations to realize the degrees of freedom of an arbitrary Mueller matrix. One significant implication of the quantum channels described here is that polarimetry can consistently be represented in a trace-preserving manner. This ensures total probability conservation for both deterministic and nondeterministic Mueller matrices, even though some information may leak to the environment about the states and transformations involved.

\subsection{Stokes operators}

The electric field of a monochromatic electromagnetic wave propagating along the $\vec{k}$ direction is given by
\begin{eqnarray}
\vec{E}(\vec{r},t)=(h\vec{h}+v\vec{v})e^{i\vec{k}\cdot\vec{r}-i\omega t}
\end{eqnarray}
where $h,v$ are two complex constants and $\vec{h},\vec{v}$ are two vectors orthogonal to the direction of propagation $\vec{k}$. The polarization properties of the field $\vec{E}$ are characterized by the Stokes parameters
\begin{eqnarray}
S_0&=&|h|^2+|v|^2\qquad
S_1\,\,=\,\,|h|^2-|v|^2\cr\cr
S_2&=&h^*v+v^* h\qquad
S_3\,\,=\,\,-i(h^*v-v^*h)
\end{eqnarray}
as usual. Only three of these parameters are independent since
\begin{eqnarray}
S_0^2&=&S_1^2+S_2^2+S_3^2\label{ClassConst}
\end{eqnarray}
The Stokes vector
\begin{eqnarray}
\vec{S}=(S_1,S_2,S_3)
\end{eqnarray}
after normalizing by $S_0$, spans the Poincar\'e sphere. For the description of stochastic light, the Stokes parameters are given as time or ensemble averages. Consequently $\vec{S}/S_0$ typically lies inside the Poincar\'e sphere. This motivates the definition of the degree of polarization
\begin{eqnarray}
p={|\vec{S}|\over S_0}
\end{eqnarray}

To develop a quantum mechanical description it is cleanest to consider the field in some finite volume, which renders the modes of the field discrete. Towards this end, imagine the electric field is contained inside a cavity (or region) of volume $V$. The field $\vec{E}$ is quantized by promoting $h$ and $v$ to operators as follows \footnote{In what follows a hat is used to denote an operator.}
\begin{eqnarray}
h&\to&\sqrt{\hbar\omega\over 2V\epsilon_0}\hat{a}_H^\dagger\qquad\qquad
v\,\,\to\,\,\sqrt{\hbar\omega\over 2V\epsilon_0}\hat{a}_V^\dagger
\end{eqnarray}
These are the usual bosonic oscillators which obey the Fock space algebra
\begin{eqnarray}
[\hat{a}_i,\hat{a}_j^\dagger]&=&\delta_{ij}\qquad [\hat{a}_i^\dagger,\hat{a}_j^\dagger]\,\,=\,\,0\,\,=\,\,[\hat{a}_i,\hat{a}_j]
\end{eqnarray}
Following \cite{goldberg2020quantum} transform to a circularly polarized basis
\begin{eqnarray}
\hat{a}_L&=&{\hat{a}_H-i\hat{a}_V\over\sqrt{2}}\qquad\qquad \hat{a}_R\,\,=\,\,{\hat{a}_H+i\hat{a}_V\over\sqrt{2}}
\end{eqnarray}
In terms of these oscillators we can define the Stokes operators as follows
\begin{eqnarray}
\hat{S}_0&=&\hat{a}^\dagger_L\hat{a}_L+\hat{a}^\dagger_R\hat{a}_R\qquad
\hat{S}_1\,\,=\,\,\hat{a}^\dagger_L\hat{a}_R+\hat{a}^\dagger_R\hat{a}_L\cr\cr
\hat{S}_2&=&-i(\hat{a}^\dagger_L\hat{a}_R-\hat{a}^\dagger_R\hat{a}_L)\qquad
\hat{S}_3\,\,=\,\,\hat{a}^\dagger_L\hat{a}_L-\hat{a}^\dagger_R\hat{a}_R\cr
&&
\end{eqnarray}
The classical Stokes parameters are recovered as expectation values of these operators. It is not possible to do better: since the Stokes operators do not commute, simultaneous eigenvectors with eigenvalue given by the classical value of the Stokes parameter simply do not exist. The Stokes operators obey the u(2) algebra
\begin{eqnarray}
[\hat{S}_\mu,\hat{S}_\nu]&=&2i(1-\delta_{\mu 0})(1-\delta_{\nu 0})\sum_{j=1}^3\epsilon_{\mu\nu j}\hat{S}_j
\end{eqnarray}
and respect the constraint
\begin{eqnarray}
\hat{S}_1^2+\hat{S}_2^2+\hat{S}_3^2&=&\hat{S}_0^2+2\hat{S}_0
\end{eqnarray}
which is the quantum analogue of (\ref{ClassConst}). It is interesting to note that since we are forced to use expectation values to connect to the classical description, there are many distinct quantum states that lead to the same set of classical Stokes parameters.

\subsection{Mueller Matrices, Jones Matrices and Quantum Channels}

It is an empirical fact that different materials linearly transform polarization properties of the incident light. The four Stokes parameters change as follows
\begin{eqnarray}
S_\mu&\to& S'_\mu\,\,=\,\,\sum^3_{\nu=0} M_{\mu\nu}S_{\nu}
\end{eqnarray}
where the $4\times 4$ matrix $M_{\mu\nu}$ is known as the Mueller matrix. Mueller matrices are broadly categorized as either depolarizing or non-depolarizing. Non-depolarizing Mueller matrices can be described using Jones matrices $J$, which represent an SL(2,$\mathbb{C}$) \footnote{The group SL(2,$\mathbb{C}$) consists of all $2\times 2$ complex matrices with determinant 1. There is a two-to-one homomorphism between SL(2,$\mathbb{C}$) and the proper orthochronous Lorentz group SO$^+$(1,3), which is the subgroup of the Lorentz group that preserves the orientation of space and the direction of time. This means that each element of SO$^+$(1,3) corresponds to two elements in SL(2,$\mathbb{C})$. Some of the manipulations below have a transparent interpretation, bearing this connection in mind.} transformation on the electric field
\begin{eqnarray}
|E\rangle \equiv \left[\begin{array}{c} h\\ v\end{array}\right]\to
\left[\begin{array}{cc} J_{11} &J_{12}\\ J_{21} &J_{22}\end{array}\right]
\left[\begin{array}{c} h\\ v\end{array}\right]\equiv J |E\rangle\end{eqnarray}
The non-depolarizing Mueller matrices do not change the degree of polarization of perfectly polarized incident light. The depolarizing Mueller matrices are described as ensemble averages of Jones matrices
\begin{eqnarray}
|E\rangle\langle E|&\to& \sum_i\lambda_i J_i |E\rangle\langle E| J_i^\dagger
\end{eqnarray}
and they reduce the degree of polarization of perfectly polarized incident light. Partially polarized light can have its degree of polarization both increase and decrease under the effect of both depolarizing and non-depolarizing optical systems. Non-depolarizing systems are also called deterministic since they are realized as pure Jones matrices, while depolarizing systems are non-deterministic because they are realized as ensemble averages of Jones matrices. This language has a natural parallel for quantum systems.

It is straight forward to derive a relationship between the Jones matrices and the Mueller matrices. In terms of the Pauli matrices (\ref{PauliMats}) and the $2\times 2$ identity matrix $\sigma_0$, it is simple to verify that \footnote{This connection simply reflects the fact that the Stokes parameters are in the 4-vector $({1\over 2},{1\over 2})$ representation of the Lorentz group. Indeed, the indices of the Pauli matrices are in the $({1\over 2},0)$ and the $(0,{1\over 2})$ and $({1\over 2},0)\otimes (0,{1\over 2})=({1\over 2},{1\over 2})$.}
\begin{eqnarray}
|E\rangle\langle E|&=&{1\over 2}\left(\sigma_0 S_0+\sigma_1 S_2+\sigma_2 S_3+\sigma_3 S_1\right)\cr\cr
&=&{1\over 2}\sum_{\mu=0}^3\sum_{\nu=0}^3\sigma_\mu A_{\mu\nu}S_\nu
\end{eqnarray}
with
\begin{eqnarray}
A=\left[\begin{array}{cccc} 1 &0 &0 &0\\0 &0 &1 &0\\0 &0 &0 &1\\0 &1 &0 &0\end{array}\right]\label{Adfnd}
\end{eqnarray}
This implies that
\begin{eqnarray}
S_\mu&=&\sum_{\nu=0}^3 A^T_{\mu\nu}{\rm Tr}(\sigma_\nu |E\rangle\langle E|)
\end{eqnarray}
where $A^T$ is the usual matrix transpose of $A$. Under the action of the Jones matrix we have
\begin{eqnarray}
S_\mu\,\,\to\,\,S'_\mu&=&\sum_{\nu=0}^3 A^T_{\mu\nu}{\rm Tr}(\sigma_\nu J|E\rangle\langle E|J^\dagger)\cr\cr
&=&{1\over 2}\sum_{\nu=0}^3\sum_{\alpha=0}^3\sum_{\beta=0}^3A^T_{\mu\nu}{\rm Tr}(\sigma_\nu J\sigma_\alpha J^\dagger) A_{\alpha\beta}S_\beta\cr\cr
&=&\sum_{\beta=0}^3M_{\mu\beta}S_\beta\label{TransMtoJ}
\end{eqnarray}
From this we can read off the relation between the Mueller matrix and the Jones matrix (for a relevant and related textbook discussion, see Section 8.3.2 of \cite{nielsen2001quantum})
\begin{eqnarray}
M_{\mu\beta}&=&{1\over 2}\sum_{\nu=0}^3\sum_{\alpha=0}^3
A^T_{\mu\nu}{\rm Tr}(\sigma_\nu J\sigma_\alpha J^\dagger) A_{\alpha\beta}
\end{eqnarray}
The density matrix for photon $B$ will appear exactly as $|E\rangle\langle E|$ appears in the above discussion. This makes it clear that the Jones matrices are the Krauss operators of the operator sum representation for the noise channel. The discussion of the Krauss operators acting on the density matrix of the state can be phrased, after using (\ref{TransMtoJ}), in terms of an affine mapping of the Stokes vectors.

Mueller matrices for common elements used to control polarization are well known and we will use these descriptions to construct noise models. These elements include
\begin{itemize}
\item[1.] {\bf Retarders:} maintain the intensity $S_0$ and the degree of polarization $p$, but they rotate the polarization vector $\vec{S}$ so that they are deterministic. The Mueller matrix is
\begin{eqnarray}
M_R=\left[\begin{array}{cc} 1 &0\cr 0^T &R\end{array}\right]\label{retarder}
\end{eqnarray}
where $R$ is a $3\times 3$ rotation matrix.
\item[2.] {\bf Diattenuators:} differentially transmit light incident with different polarization directions. They take perfectly polarized light to perfectly polarized light at a reduced intensity so that they are also deterministic. The Mueller matrix is ($0< q,r < 1$)
\begin{eqnarray}
M_D=\left[\begin{array}{cccc}{q+r\over 2} &{q-r\over 2} &0 &0\\ {q-r\over 2} &{q+r\over 2} &0 &0\\ 0 &0 &\sqrt{qr} &0\\ 0 &0 &0 &\sqrt{qr}\end{array}\right]\label{diattenuator}
\end{eqnarray}
\item[3.] {\bf Depolarizers:} maintain the total intensity $S_0$ but reduce the degree of polarization $p$ of perfectly polarized light. The Mueller matrix is
\begin{eqnarray}
M_d =\left[\begin{array}{cc} 1 &0\\ 0^T &m\end{array}\right]\label{depolarizer}
\end{eqnarray}
where $m$ is a $3\times 3$ symmetric matrix with eigenvalues between $-1$ and 1. Setting $r=0$ and $q=1$ in $M_D$ gives a Mueller matrix for the linear polarizer.
\item[4.] {\bf Arbitrary Mueller Matrices:} can be decomposed into the sequential application of a diattenuator, a retarder, and a depolarizer $M=M_dM_DM_R$.
\end{itemize}

It is possible to derive the quantum operation that corresponds to any given Mueller matrix. General quantum channels are represented by completely positive and trace preserving (CPTP) maps acting on the density matrix $\rho=|E\rangle\langle E|$
\begin{eqnarray}
{\cal E}(\hat\rho)&=&\sum_l \hat{K}_l\hat{\rho}\hat{K}_l^\dagger\qquad\qquad
\sum_l \hat{K}_l^\dagger\hat{K}_l\,\,=\,\,{\bf 1}
\end{eqnarray}
This is called the operator sum representation of the channel and the operators $\hat{K}_l$ are known as Krauss operators. Composite quantum channels are produced as products or sums of other quantum channels. The application of Mueller matrix $M_1$ followed by $M_2$ constructs the composite channel
\begin{eqnarray}
{\cal E}_{M_2M_1} (\hat{\rho})&=&{\cal E}_{M_2} \left({\cal E}_{M_1}(\hat{\rho})\right)
\end{eqnarray}
while a linear combination of matrices $M_1$ and $M_2$ constructs the channel
\begin{eqnarray}
{\cal E}_{p_1M_1+p_2M_2}(\hat{\rho})&=&p_1\, {\cal E}_{M_1}(\hat{\rho})+p_2\, {\cal E}_{M_2}(\hat{\rho})
\end{eqnarray}
Synthesizing more complicated quantum channels from simpler building blocks reduces the problem of constructing the channel for an arbitrary Mueller matrix to that of constructing channels for retarders, diattentuators and depolarizers. This construction is straight forward as we now explain.

{\vskip 0.5cm}

\noindent
{\bf Quantum Channels for Retarders:} To realize the retarder, (\ref{retarder}) instructs us to construct a rotation. The Mueller matrix in (\ref{retarder}) acts on the Stokes parameters $S_1$, $S_2$ and $S_3$. We need to construct a Jones matrix, which will act on the state of our polarization photon (photon $B$). The relevant argument is given in (\ref{TransMtoJ}). Concretely ($i,j=1,2,3$ and $A_{ij}$ is the lower right 3$\times$3 block of matrix $A$ defined in (\ref{Adfnd}).)
\begin{eqnarray}
S_i&=&\sum_{j=1}^3 A^T_{ij}{\rm Tr}(\sigma_j J|E\rangle\langle E|J^\dagger)\cr\cr
&=&{1\over 2}\sum_{j=1}^3\sum_{k=1}^3\sum_{l=1}^3A^T_{ij}{\rm Tr}(\sigma_j J\sigma_k J^\dagger) A_{kl}S_l\cr\cr
&=&\sum_{l=1}^3 R_{ik}S_l
\end{eqnarray}
so that we read off\footnote{It is often stated that many Jones matrices correspond to one Mueller matrix because $R_{ij}$ is independent of the overall phase of $J$. This independence of the phase is already present at the level of the density matrix and we never distinguish states that only differ by an overall phase. Thus, at quantum level the phase of the Jones matrix (which for us is a Krauss operator) is not physical.}
\begin{eqnarray}
R_{ik}&=&{1\over 2}\sum_{j=1}^3\sum_{k=1}^3\sum_{l=1}^3A^T_{ij}{\rm Tr}(\sigma_j J\sigma_k J^\dagger) A_{kl}
\end{eqnarray}
This relation can be used to verify that the rotation, written using Euler angles
\begin{widetext}
\begin{eqnarray}
R^T&=&\left[
\begin{array}{ccc}
\cos\theta\cos\varphi\cos\psi-\sin\varphi\sin\psi &\cos\psi\sin\varphi+\cos\theta\cos\varphi\sin\psi &-\cos\varphi\sin\theta\\
-\cos\theta\cos\psi\sin\varphi-\cos\varphi\sin\psi &\cos\psi\cos\varphi-\cos\theta\sin\varphi\sin\psi &\sin\varphi\sin\theta\\
\cos\psi\sin\theta &\sin\theta\sin\psi &\cos\theta
\end{array}
\right]\cr\cr
&&
\end{eqnarray}
\end{widetext}
corresponds to the Jones matrix\footnote{Familiar group theory of SO$^+$(1,3) is again apparent. The rotation $R_{ik}$ rotates the spatial components of the 4-vector in $({1\over 2},{1\over 2})$ while the Jones matrix is the same transformation in the $({1\over 2},0)$ representation.}
\begin{eqnarray}
J_R(\theta,\varphi,\psi)&=&\left[
\begin{array}{cc}
e^{-{i\over 2}(\varphi+\psi)}\cos{\theta\over 2} &e^{-{i\over 2}(\varphi-\psi)}\sin{\theta\over 2}\\
-e^{{i\over 2}(\varphi-\psi)}\sin{\theta\over 2} &e^{{i\over 2}(\varphi+\psi)}\cos{\theta\over 2}
\end{array}
\right]\label{JonesRetarder}
\end{eqnarray}
This Jones matrix is the Krauss operator defining the quantum channel for a retarder. This channel is deterministic, so there is a single Krauss matrix.

{\vskip 0.5cm}

\noindent
{\bf Quantum Channels for Diattentuators:} The different components of the electric field are differentially transmitted. For example, the transformation $h\to\sqrt{q}h$ and $v\to\sqrt{r}v$ corresponds to transmission probabilities $q$ and $r$ for horizontally and vertically polarized light. This transformation is implemented by the diattenuator Mueller matrix $M_d$ given in (\ref{diattenuator}). A natural guess is that $\hat{a}_H\to\sqrt{q}\hat{a}_H$ and $\hat{a}_V\to\sqrt{r}\hat{a}_V$ provides the channel for a diattenuator. This simple transformation does not preserve the oscillator commutation relations. The paper \cite{goldberg2020quantum} constructs the correct transformation by moving to an enlarged Hilbert space and then tracing out the auxilliary modes. The resulting Jones matrix is given by
\begin{widetext}
\begin{eqnarray}
J_d(\theta,\psi,r,q)&=&{1\over 2}\left[
\begin{array}{cc}
\sqrt{q}+\sqrt{r}+(\sqrt{q}-\sqrt{r})\cos\theta &e^{i\psi}(\sqrt{q}-\sqrt{r})\sin\theta\\
e^{-i\psi}(\sqrt{q}-\sqrt{r})\sin\theta &\sqrt{q}+\sqrt{r}-(\sqrt{q}-\sqrt{r})\cos\theta
\end{array}
\right]\label{JonesDiattenuator}
\end{eqnarray}
\end{widetext}
This provides the Krauss matrix needed to define the diattenuation channel. To achieve diattenuation along a different direction, apply a rotation followed by linear diattenuation and then the inverse of the original rotation. The rotation is parameterized by only two angles because only the direction of the diattenuation axis needs to be varied. Consequently, the four parameters of a general Mueller matrix representing diattenuation are captured by $q$, $r$, and the two angles specifying the diattenuation axis. Since this channel is deterministic, it involves a single Krauss matrix.

{\vskip 0.5cm}

\noindent
{\bf Quantum Channels for Depolarizers:} Number conserving quantum operations account for the seven parameters of non-depolarizing Mueller matrices. The remaining nine free parameters are described by quantum operations that don't conserve photon number \cite{goldberg2020quantum}. Any map from the space of density matrices back to the space of density matrices must be completely positive and trace-preserving (CPTP). The CPTP requirement enforces the known constraints for Mueller matrices, which were derived classically in \cite{gamel2011causality}. This strongly suggests a quantum-mechanical origin for the constraint that depolarizing Mueller matrices are formed from positive combinations of non-depolarizing Mueller matrices \cite{simon2010nonquantum}. CPTP maps that transform Stokes operators into linear combinations of Stokes operators always allow a decomposition into convex combinations of CPTP maps corresponding to non-depolarizing Mueller matrices (since we study maps that do not mix different photon-number subspaces, the results of \cite{gamel2011causality} validate this). In the single-photon subspace, the CPTP maps precisely match the classical constraint requiring convex combinations of non-depolarizing Mueller matrices. Therefore, convex combinations of the described channels are sufficient to represent all classical Mueller matrices. If the Krauss operators $\{\hat{K}^{(i)}_l\}$ correspond to Mueller matrix $M(i)$, then the quantum channel
\begin{eqnarray}
{\cal E}(\hat{\rho})&=&\sum_i p_i\sum_l \hat{K}(i)_l\hat{\rho}\hat{K}(i)^\dagger_l
\end{eqnarray}
corresponds to the Mueller matrix
\begin{eqnarray}
M&=&\sum_i \,p_i\,M(i)
\end{eqnarray}
It is a proven result that no more than four weights $p_i$ are ever needed \cite{cloude1986group}. 

{\vskip 0.5cm}

This provides a complete description of the channels needed for a description of the polarization noise affecting photon $B$.

\section{Topological noise rejection}\label{TNR}

A central message of this study is that the Skyrmion number exhibits remarkable robustness against noise. In this section, we will explain why this robustness arises from encoding information into topology, which is sensitive only to the global features of the wave function. The fundamental concept is that noise primarily affects the local features of the wave function, so encoding information into its topology protects the information from the effects of noise.

Topology is a systematic approach to characterizing quantities that are insensitive to smooth deformation. In our problem, this is the statement that the Skyrmion number is unchanged by a change of coordinates $\vec{r}_A$ for photon A. To demonstrate this, start from the formula for the Skyrmion number
\begin{equation}
    N = \frac{1}{4\pi}\int\limits_{-\infty}^{\infty}\!\int\limits_{-\infty}^{\infty} \epsilon_{pqr} S_p \frac{\partial S_q}{\partial x} \frac{\partial S_r}{\partial y}  dx dy  \label{eqn:TopInvSkyNum}\,
\end{equation}
and consider a smooth change of coordinates from $x,y$ to $x'\,=\,x'(x,y)$ and $y'\,=\,y'(x,y)$. After some simple manipulations, we find the following transformation of the integrand 
%
%\begin{eqnarray}
%    \epsilon_{zqr} S_p \frac{\partial S_q}{\partial x} \frac{\partial S_r}{\partial y}&=& \epsilon_{zqr} S_p \left(\frac{\partial x'}{\partial x}\frac{\partial S_q}{\partial x'}+\frac{\partial y'}{\partial x}\frac{\partial S_q}{\partial y'}\right)
%\left(\frac{\partial x'}{\partial y}\frac{\partial S_r}{\partial x'}+\frac{\partial y'}{\partial y}\frac{\partial S_r}{\partial y'}\right)\cr\cr\cr
%&=& \epsilon_{zqr} S_p \left(\frac{\partial x'}{\partial x}\frac{\partial x'}{\partial y}\frac{\partial S_q}{\partial x'}\frac{\partial S_r}{\partial x'}+\frac{\partial x'}{\partial x}\frac{\partial y'}{\partial y}\frac{\partial S_q}{\partial x'}\frac{\partial S_r}{\partial y'}\right.\cr\cr
%&&\qquad\qquad\qquad\left.+\frac{\partial y'}{\partial x}\frac{\partial x'}{\partial y}\frac{\partial S_q}{\partial y'}\frac{\partial S_r}{\partial x'}+\frac{\partial y'}{\partial x}\frac{\partial y'}{\partial y}\frac{\partial S_q}{\partial y'}\frac{\partial S_r}{\partial y'}\right)
%\end{eqnarray}
%%
%The first and fourth terms above are sums of an expression that is symmetric under interchange of $q$ and $r$, times $\epsilon_{zqr}$ which is antisymmetric under the interchange of $q$ and $r$, so these terms vanish. In the third term relabel the indices $q\leftrightarrow r$ to obtain
%%
\begin{eqnarray}
\epsilon_{pqr} S_p \frac{\partial S_q}{\partial x} \frac{\partial S_r}{\partial y} &=& \left(\frac{\partial x'}{\partial x}\frac{\partial y'}{\partial y}-\frac{\partial y'}{\partial x}\frac{\partial x'}{\partial y}\right) \epsilon_{pqr} S_p \frac{\partial S_q}{\partial x'}\frac{\partial S_r}{\partial y'}\nonumber
\end{eqnarray}
The integration measure transforms as $dx dy =J dx' dy'$ where the Jacobian is given by
\begin{eqnarray}
J&=&\left(\frac{\partial x}{\partial x'}\frac{\partial y}{\partial y'}-\frac{\partial y}{\partial x'}\frac{\partial x}{\partial y'}\right)\,\,=\,\,\left(\frac{\partial x'}{\partial x}\frac{\partial y'}{\partial y}-\frac{\partial y'}{\partial x}\frac{\partial x'}{\partial y}\right)^{-1}\nonumber
\end{eqnarray}
Putting these transformation rules together it follows that
\begin{eqnarray}
    N &=& \frac{1}{4\pi}\int\limits_{-\infty}^{\infty}\!\int\limits_{-\infty}^{\infty} 
\frac{1}{2} \epsilon_{pqr} S_p \frac{\partial S_q}{\partial x} \frac{\partial S_r}{\partial y}  dx dy\cr\cr
      &=& \frac{1}{4\pi}\int\limits_{-\infty}^{\infty}\!\int\limits_{-\infty}^{\infty} \frac{1}{2} \epsilon_{pqr} S_p \frac{\partial S_q}{\partial x'} \frac{\partial S_r}{\partial y'} dx' dy' 
\end{eqnarray}
proving that the Skyrmion number is independent of the choice of coordinates used for the calculation. This demonstrates that the Skyrmion number remains invariant under smooth deformations, as any such deformation is effectively equivalent to a carefully selected change of coordinates. Consequently, the Skyrmion number is perfectly robust against all sources of noise that can be modelled as a smooth deformation.

\begin{figure*}[t!]
\includegraphics[width=1.0\linewidth]{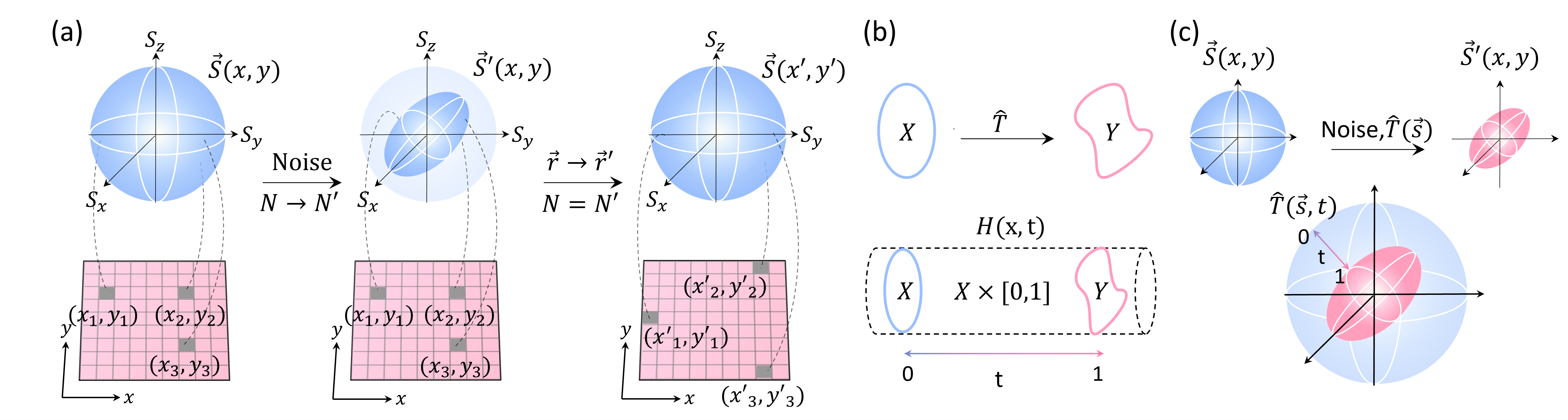}
\caption{\textbf{The origin of topologically protected wave functions.} (a) The map $\vec{S}(x,y)$ (left panel) derived from the Skyrmion wave function, is a map from the plane to the sphere $R^2 \to S^2$. The map is distorted when the wave function is exposed to noise (middle panel), $\vec{S}(x,y) \to \vec{S}'(x,y)$. However, (right panel) if this distortion can be inverted with a smooth coordinate transformation $(x,y) \to (x',y')$, then the noise will not alter the topology of the state $N\to N'=N$ realizing topological protection. (b) To demonstrate that transformation $\hat{T}$ is a smooth deformation of topological space $X$ into topological space $Y$, we construct a smooth homotopy $H(t,x)$ with $t\in [0,1]$ that enjoys the properties $H(0,x) = \textbf{1}$ and $H(1,x) = \hat{T}$. (c) The noise channel distorts the initial map as $\hat{T}(\vec{S}(x,y))=\vec{S}'(x,y)$. If this channel can be smoothly parametrized with parameter $t$ such that $\hat{T}(0,\vec{S}(x,y))=\vec{S}(x,y)$ is the identity and at $t=1$, $\hat{T}(1,\vec{S}(x,y))=\vec{S}'(x,y)$ then the noise channel induces a smooth deformation of the map defined by the Skyrmion wave function proving topological protection of the Skyrmion topology.}
\label{fig:coordtransform}
\end{figure*}

This concept is illustrated in Fig.~\ref{fig:coordtransform} (a), where noise distorts the original Skyrmion map, $\vec{S}(x,y)$, into a noisy map $\vec{S'}(x,y)$, potentially altering the topology from $N$ to $N'$. Since we are working with maps to unit spheres, $S^2$, we perform a normalization to ensure that $\vec{S'}(x,y)\cdot\vec{S'}(x,y) = 1$. After this normalization, we check whether a coordinate transformation (smooth deformation), $(x,y) \to (x',y')$, exists such that $\vec{S'}(x,y) \to \vec{S}(x',y')$. If such a transformation exists, then the noise is a smooth deformation, leaving the topology unchanged.

%Why does the freedom to change coordinates lead to resilience against noise? Consider a single spatial dimension for simplicity. Noise transforms the wave function $\psi(x)$ into $\psi'(x)$. If the added noise is modest, the relation $\psi'(x)=\psi(x')$ will always define a coordinate change from $x$ to a new coordinate $x'$. Consider the topological invariant associated with the noisy wave function. Since topological invariants are unchanged by coordinate transformations, we can freely perform the transformation $x\to x'$ as defined. This transformation replaces $\psi'(x)$ with $\psi(x')$, effectively returning us to the noise free wave function in the new coordinates. Thus the topological invariant associated to the noisy wave function is equal to that associated with the noise free wave function, which illustrates the topological noise rejection mechanism.

This intuitive argument can be formalized in our setting using the concept of homotopy. In topology, two continuous functions between topological spaces are called homotopic if one can be ``continuously deformed'' into the other. The deformation that connects the two functions is known as a homotopy. More formally, a homotopy between two continuous functions, $f(x)$ and $g(x)$, from a topological space $X$ (represented as a loop in Fig.~\ref{fig:coordtransform} (b)) to a topological space $Y$ (shown as a distorted loop), is a continuous function $H: X \times [0,1] \to Y$ from the product of space $X$ and the unit interval $[0,1]$ to $Y$. This function satisfies the conditions $H(x,0) = f(x)$ and $H(x,1) = g(x)$ for all $x \in X$. Since topological invariants, such as the Skyrmion number, are preserved under homotopy, demonstrating robustness of our discrete signal against noise (denoted as $\hat{T}(\vec{s},t)$ in Fig.~\ref{fig:coordtransform} (c)) requires showing the existence of a homotopy between the noisy function, $\hat{T}(\vec{s},1)$, and the noise-free function, $\hat{T}(\vec{s},0)$. In prior classical studies of vectorial light, this homotopy has been assumed rather than proven \cite{wang2024topological}. By establishing homotopies between the noisy and noise-free density matrices, we confirm the existence of a smooth deformation from the original to the noisy wave function, thus generalizing prior classical analysis \cite{wang2024topological}.

\section{Topological protection against non-depolarizing channels}\label{homotopy}

The non-depolarizing channels are described by pure Mueller matrices and they correspond to retarders and diattentuators. For pure Mueller matrices there is a single term in the quantum channel. Topological invariance is established with a homotopy which interpolates between the original density matrix, and the density matrix output of the quantum channel. For these channels the Krauss operator is simply given by the Jones matrix associated to the Mueller matrix. The noise model of Section \ref{NoiseModel} allows the parameters of the Jones matrix to depend on position $\vec{r}_A$. Denote the space of $\vec{r}_A=(x,y)$ points $\mathbb{R}_A^2$ where $A$ is for photon $A$. Denote the space of polarization states, the Poincare sphere, as S$^2_B$ where $B$ is for photon $B$.

The formula for the Stokes parameters (\ref{StokesP}) uses the diagonal matrix element of the density matrix, with respect to the position space basis of photon $A$
\begin{eqnarray}
\rho_B\equiv \langle\vec{r}_A|\psi\rangle\langle\psi|\vec{r}_A\rangle=\langle\vec{r}_A|\rho|\vec{r}_A\rangle
\end{eqnarray}
$\rho_B$ is a 2$\times$2 matrix and it is a density matrix in the sense that it is positive
\begin{eqnarray}
\langle v|\rho_B|v\rangle\ge 0\qquad\qquad\forall |v\rangle
\end{eqnarray}
Further, it is natural to normalize $\rho_B$ at each point $\vec{r}_A$ since Stokes parameters computed with the normalized $\rho_B$ will be correctly normalized.

\subsection{Retarder Homotopy}

A retarder is described using the Jones matrix given in (\ref{JonesRetarder}). Notice that this Jones matrix is an element of SU(2). The angles $\varphi,\psi,\theta$ characterizing the retarder are all functions from $\mathbb{R}^2_A$ to the interval $[0,2\pi]$. We now define the homotopy $H_R: \mathbb{R}^2_A\times [0,1]\to $SU$(2)$ as follows
\begin{eqnarray}
H_R(t,x,y)&=&\left[
\begin{array}{cc}
e^{-{i\over 2}(\varphi_t+\psi_t)}\cos{\theta_t\over 2} &e^{-{i\over 2}(\varphi_t-\psi_t)}\sin{\theta_t\over 2}\\
-e^{{i\over 2}(\varphi_t-\psi_t)}\sin{\theta_t\over 2} &e^{{i\over 2}(\varphi_t+\psi_t)}\cos{\theta_t\over 2}
\end{array}
\right]\cr
&&
\end{eqnarray}
where $0\le t\le 1$ and
\begin{eqnarray}
\theta_t&=& t\theta (x,y)\qquad \varphi_t\,\,=\,\, t\varphi (x,y)\qquad \psi_t\,\,=\,\, t\psi (x,y)\cr
&&\label{anglefunctions}
\end{eqnarray}
Notice that $H_R(0,x,y)={\bf 1}$ and $H_R(1,x,y)=J_R(\theta,\varphi,\psi)$ where ${\bf 1}$ is the 2$\times$2 identity matrix and $J_R(\theta,\varphi,\psi)$ is the Jones matrix defined in (\ref{JonesRetarder}). The $\sin(\cdot)$, $\cos(\cdot)$ and $e^{(\cdot)}$ functions are all infinitely differentiable, so as long as the functions defined in (\ref{anglefunctions}) are smooth functions of $(x,y)$, we obtain a smooth homotopy from the undeformed state
\begin{eqnarray} 
H_R(0,x,y)\rho_B H_R(0,x,y)&=&\rho_B
\end{eqnarray} 
to the noisy state 
\begin{eqnarray}
H_R(1,x,y)\rho_B H_R(1,x,y)&=&J_R(\theta,\varphi,\psi)\rho_BJ_R(\theta,\varphi,\psi)^\dagger\cr
&&
\end{eqnarray}
The fact that the Krauss operator, which is given by a Jones matrix representing a rotation, can be smoothly deformed to the identity, follows immediately simply from the fact that SU(2) is a connected Lie group. In a connected Lie group, every element can be smoothly deformed to the identity. The continuous deformation to the identity is the homotopy establishing that the Skyrmion number is unchanged by the action of a retarder. Further, the existence of this smooth mapping demonstrates that this invariance is a consequence of topological noise rejection against the effects of retarders.

It is worth noting that we can give an independent analytic demonstration of the independence of the Skyrmion number against the effects of the retarder, when the parameters of the retarder are constant. The Mueller matrix from the retarder, given in equation (\ref{retarder}), show that the retarder simply rotates the spatial components of the Stokes vector: $S_i\to \sum_j R_{ij}S_j$. The Skyrmion number density therefore becomes %
\begin{eqnarray}
\sum_{ijk}\epsilon_{ijk}S_i\partial_x S_j\partial_y S_k&\to&\sum_{ijk}\sum_{lmn}\epsilon_{ijk}R_{il}R_{jm}R_{kn}S_l\partial_x S_m\partial_y S_n\nonumber
\end{eqnarray}
Now, use the identity
\begin{eqnarray}
\sum_{ijk}\epsilon_{ijk}R_{il}R_{jm}R_{kn}&=&\det (R)\epsilon_{lmn}\,\,=\,\,\epsilon_{lmn}
\end{eqnarray}
to conclude that the Skyrmion number density is invariant under the action of the retarder.

\begin{figure}[t!]
\includegraphics[width=1.0\linewidth]{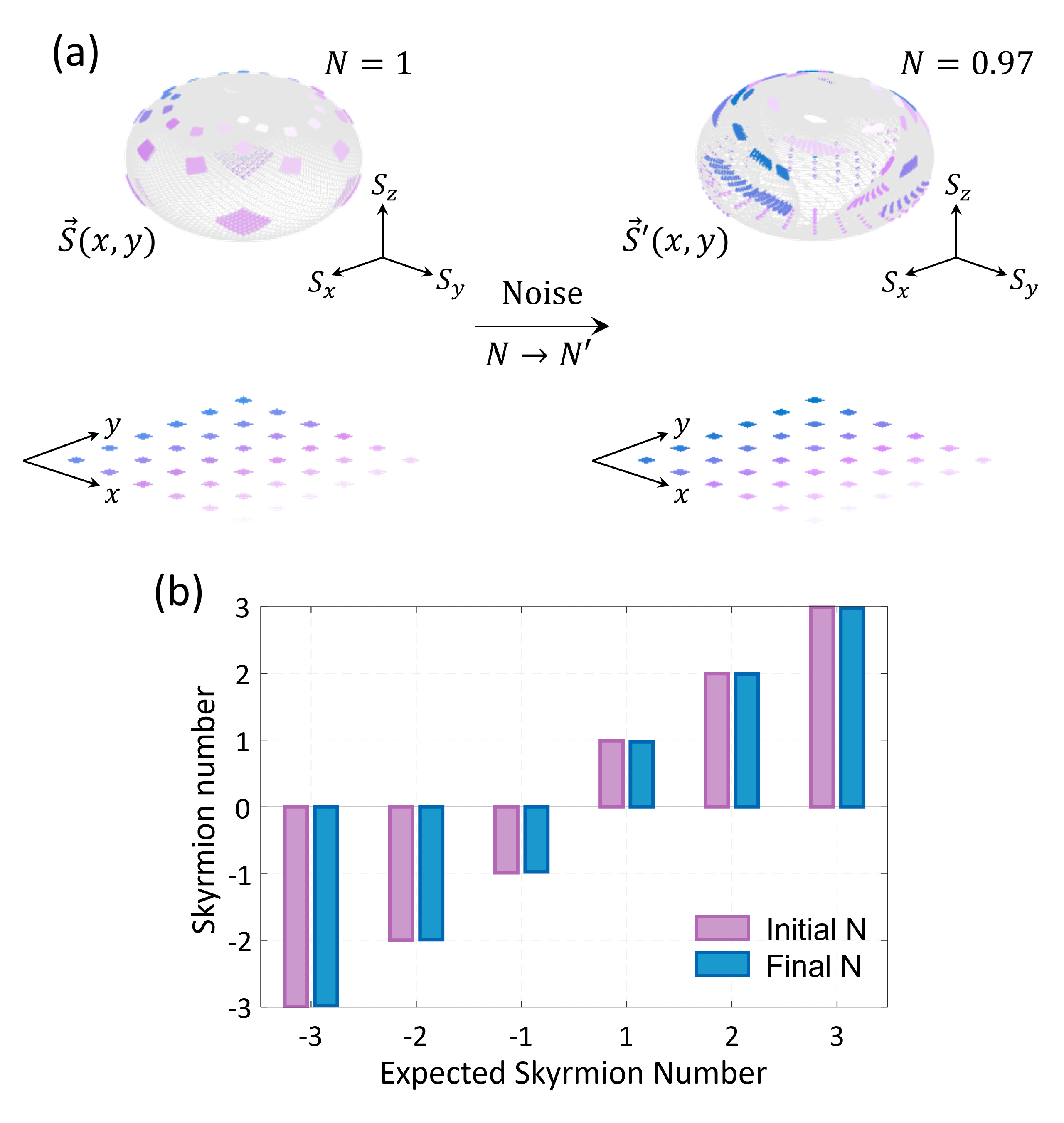}
\caption{\textbf{Skyrmion resilience against Retarder channel.} (a) Skyrmion map for $N=1$ shown before (left panel) and after (right panel) passing through a spatially varying retarder channel. Although the mapping is altered, $\vec{S}(x,y) \to \vec{S}'(x,y)$, the net affect is a shifting of points on the sphere without changing the wrapping number. Indeed, the calculated Skyrmion number for the noisy signal is $N \approx 0.97$. To illustrate the movement of points under the influence of noise, we have color-coded regions of interest on the plane and the sphere. These points can be returned to their original configuration by transforming points in the plane, i.e., $\vec{S}(x,y) \to \vec{S}'(x,y) \to \vec{S}(x',y')$. This explains why the Skyrmion number is robust. (b) Numerically calculated Skyrmion number before (initial) and after (final) passing the state through the noisy channel, compared to the expected Skyrmion number for various topologies $N \in \{-3,-2,-1,1,2,3\}$. The plot demonstrates the invariance of the topology under the influence of the spatially varying retarder channel.}
\label{fig:Retarder}
\end{figure}

The analysis in this subsection demonstrates that, assuming the functions defined in (\ref{anglefunctions}) are smooth functions of $(x,y)$, we can construct a smooth homotopy from the undeformed state to the noisy state. It is straight forward to test this conclusion numerically with a few conveniently chosen smooth functions. For this exercise it proves useful to move to polar coordinates $x=\rho\cos\phi$ and $y=\rho\sin\phi$. Assume that the angle parameters of the retarder have the following coordinate dependence
\begin{eqnarray}
\theta&=& \theta_0 e^{-\beta_1\rho^2}\cos (n_1\phi)\cr\cr
\varphi&=&\varphi_0 e^{-\beta_2\rho^2}\cos (n_2\phi)\cr\cr
\psi&=& \psi_0 e^{-\beta_3\rho^2}\cos (n_3\phi)
\end{eqnarray}
Using the Jones matrix (\ref{JonesRetarder}) as the Krauss matrix for a quantum channel we can evaluate the Skyrmion number after the channel is applied. When the angle parameters $\theta$, $\varphi$ and $\psi$ vanish the Jones matrix evaluates to the identity. The constant parameters $\beta_i$ for $i=1,2,3$ control how rapidly the angle parameters fall to zero with increasing $\rho$ so that the retarder acts in some local region centred on the origin $\rho=0$. The constant parameters $\beta_i$ for $i=1,2,3$ control the spatial variation of the retarder parameters when $\rho$ is fixed and $\phi$ is varied. Finally the constant parameters $\theta_0$, $\varphi_0$ and $\psi_0$ set the magnitude of the retarder angles. In the current and proceeding sections we will numerically evaluate the affect of different noise types on the Skyrmion number of a given state. We note that deviations from the initial Skyrmion number are expected due to two main factors: (1) the discretization of the component fields, $u_0^{l_i}$, of $\rho_B$ and (2) the truncation of the integrand in equation (\ref{SkNumb}). Collectively, these two numerical factors lead to calculated Skyrmion numbers with a slightly lower magnitude than the expected value. However, both factors can be mitigated by increasing the local pixel density (the number of pixels used to describe the field) as well as increasing the region of interest with respect to the Gaussian width, $\omega_0$, of the fields at the cost of computational resources. Numerical results for various Skyrmion topologies passed through a spatially varying retarder channel are shown in Fig.~\ref{fig:Retarder}.  In Fig.~\ref{fig:Retarder} (a) the initial map $\vec{S}(x,y)$ with a Skyrmion number of $N=1$ is shown as a covered sphere with regions of interest (ROIs) highlighted and correlated by colour to the corresponding  ROIs in the plane. After passing the state through a spatially varying retarder, $\vec{S}(x,y)$ is transformed to $\vec{S}'(x,y)$. The only effect of this transformation is a rotation of the Stokes vectors as indicated by the shift in position of the ROIs on the sphere. Importantly, the full coverage of the sphere is maintained, thus leaving the Skyrmion number unaltered, as shown in Fig.~\ref{fig:Retarder} (b), reporting numerical results for states possessing Skyrmion numbers $N= \{ -3,-2,-1,1,2,3\}$ before (initial) and after (final) they were passed through the spatially varying retarder channel are shown. From these results it is clear that the initial state's topology is invariant under the action of the spatially varying retarder, thus confirming that the Skyrmion number is invariant for the above $\rho,\phi$ dependence of the retarder parameters, in agreement with the homotopy construction.

\subsection{Diattenuator Homotopy}

The channel corresponding to a diattenuator has Krauss matrix given by the Jones matrix in (\ref{JonesDiattenuator}). The channel parameters are angles $\theta,\psi$ which are functions from $\mathbb{R}^2_A$ to the interval $[0,2\pi)$, as well as the non-negative parameters $q,r$ which are functions from $\mathbb{R}^2_A$ to $(0,1)$. We now define the homotopy $H_d: \mathbb{R}^2_A\times [0,1]\to $SU$(2)$ as follows
\begin{widetext}
\begin{eqnarray}
H_d(t,x,y)&=&
{1\over 2}\left[
\begin{array}{cc}
\sqrt{q_t}+\sqrt{r_t}+(\sqrt{q_t}-\sqrt{r_t})\cos\theta_t &e^{i\psi_t}(\sqrt{q_t}-\sqrt{r_t})\sin\theta_t\\
e^{-i\psi_t}(\sqrt{q_t}-\sqrt{r_t})\sin\theta_t &\sqrt{q_t}+\sqrt{r_t}-(\sqrt{q_t}-\sqrt{r_t})\cos\theta_t
\end{array}
\right]
\end{eqnarray}
\end{widetext}
where $0\le t\le 1$ and
\begin{eqnarray}
\theta_t&=& t\theta (x,y)\qquad\qquad\qquad  \psi_t\,\,=\,\, t\psi (x,y)\cr\cr
q_t&=&(1-t)+tq\qquad\qquad r_t=(1-t)+tr\label{difunctions}
\end{eqnarray}
This homotopy obeys $H_d(0,x,y)={\bf 1}$ and $H_d(1,x,y)=J_d(\theta,\psi,r,q)$ where ${\bf 1}$ is the 2$\times$2 identity matrix and $J_d(\theta,\psi,r,q)$ is the Jones matrix defined in (\ref{JonesDiattenuator}). The $\sin(\cdot)$, $\cos(\cdot)$ and $e^{(\cdot)}$ functions are all infinitely differentiable, so as long as the functions defined in (\ref{difunctions}) are smooth functions of $(x,y)$. The $\sqrt{(\cdot)}$ functions have square root branch points when their argument vanishes, so they are not analytic (i.e., smooth) at these points. It is however clear that this singularity is only reached for $t=(1-r)^{-1}$ or for $t=(1-q)^{-1}$. Since both $r$ and $q$ are less than 1, we never reach these singularities. Thus, we obtain a smooth homotopy from the undeformed state
\begin{eqnarray} 
H_d(0,x,y)\rho_B H_d(0,x,y)=\rho_B
\end{eqnarray} 
to the noisy state 
\begin{eqnarray}
H_d(1,x,y)\rho_B H_d(0,x,y)&=&J_d(\theta,\psi,r,q)\rho_BJ_d(\theta,\psi,r,q)^\dagger\cr
&&
\end{eqnarray}
This proves that the Skyrmion number is resilient against the effects of a diattenuator and that topological noise rejection is the mechanism behind this resilience.

\begin{figure}[t!]
\includegraphics[width=1.0\linewidth]{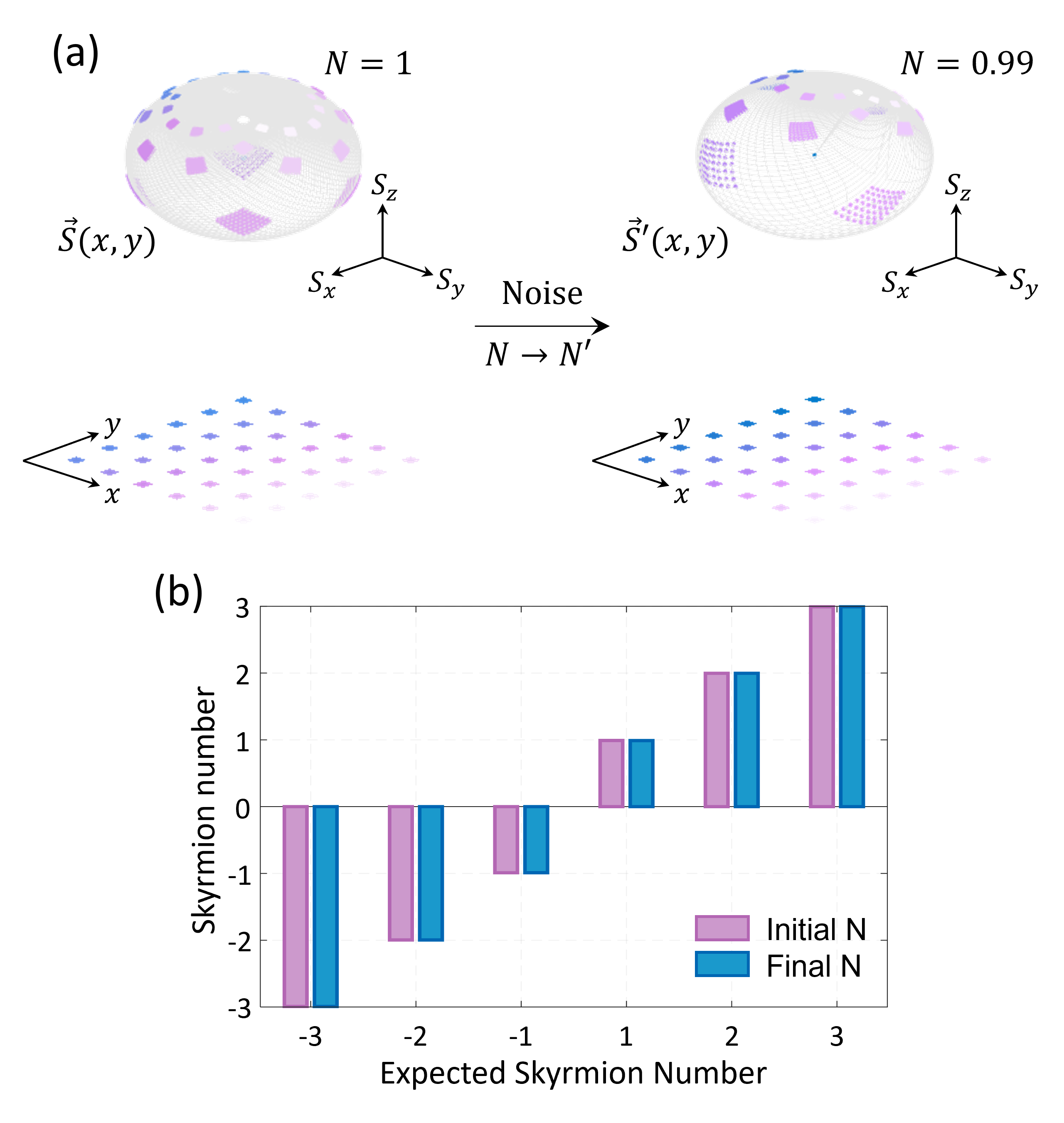}
\caption{\textbf{Skyrmion Resilience to Diattenuator Channel.} (a) Skyrmion map for $N=1$ before (left panel) and after (right panel) passing through a spatially varying diattenuator channel. While the mapping is altered $\vec{S}(x,y) \to \vec{S}'(x,y)$, points on the sphere merely shift without changing the wrapping number, leaving the Skyrmion number intact at $N\approx0.99$. This is highlighted by color-matching correlated regions of interest on the plane with those on the sphere, illustrating that the shifts on the sphere can be compensated for using a transformation of the plane, i.e., $\vec{S}(x,y) \to \vec{S}'(x,y) \to \vec{S}(x',y')$. (b) Numerically calculated Skyrmion numbers (before and after passing the state through the noisy channel) compared to expected Skyrmion numbers for various topologies, $N \in \{-3, -2, -1, 1, 2, 3\}$. The results show clear invariance of the topology to the spatially varying diattenuator channel.}
\label{fig:Diattenuator}
\end{figure}

We have argued that, under the assumption that the functions defined in (\ref{difunctions}) are smooth functions of $(x,y)$, we obtain a smooth homotopy from the undeformed state to the noisy state. We can again test this conclusion numerically, with conveniently chosen smooth functions for the channel parameters. The functions we use are given by
\begin{eqnarray}
\theta&=& \theta_0 e^{-\beta_1\rho^2}\cos (n_1\phi)\cr\cr
\psi&=& \psi_0 e^{-\beta_2\rho^2}\cos (n_2\phi)\cr\cr
q&=&(1-e^{-\beta_3\rho^2}\cos (n_3\phi))\cr\cr
r&=&(1-e^{-\beta_4\rho^2}\cos (n_4\phi))
\end{eqnarray}
As $\rho\to\infty$ the channel parameters assume the limit values $\theta=0=\psi$ and $q=1=r$. At these limiting values the Jones matrix (\ref{JonesDiattenuator}) reduces to the two dimensional unit matrix and the channel acts trivially. The constant parameters $\beta_i$ for $i=1,2,3,4$ control how rapidly these limiting values are attained as $\rho$ increases. Thus, with this choice of functions the channel again acts non-trivially is a finite region centred on $\rho=0$. The constant parameters $n_i$ for $i=1,2,3,4$ control the spatial variation of the channel parameters at fixed $\rho$ as $\phi$ is varied. Finally the constant parameters $\theta_0$ and $\psi_0$ set the magnitude of the two angles defining the diattenuator. Using the Jones matrix (\ref{JonesDiattenuator}) as the Krauss matrix for a quantum channel, with the above spatially varying channel parameters it is simple to evaluate the Skyrmion number after the channel acts. Numerical results for various Skyrmion topologies after the action of a spatially varying diattenuator channel are presented in Fig.~\ref{fig:Diattenuator}. In Fig.~\ref{fig:Diattenuator} (a), the initial map $\vec{S}(x,y)$ with a Skyrmion number of $N=1$ is transformed into $\vec{S}'(x,y)$. This transformation induces a rotation of the Stokes vectors, as shown by the shift in the position of the ROIs on the sphere. Despite this shift, the full coverage of the sphere is preserved, leaving the Skyrmion number virtually unchanged, with $N\approx0.99$. Fig.~\ref{fig:Diattenuator} (b) shows the numerical results for states with Skyrmion numbers $N= \{-3,-2,-1,1,2,3\}$ before (initial) and after (final) passing through the spatially varying diattenuator channel. These results demonstrate that the topology of the initial state is unaltered, confirming that the Skyrmion number is invariant under the specific $\rho,\phi$ dependence chosen for the diattenuator parameters, in agreement with the homotopy construction.

%Numerically we find that the Skyrmion number is invariant for the above $\rho,\phi$ dependence of the diattenuator parameters, exactly as the homotopy construction predicts.

\section{Depolarizing Channels}\label{nondepol}

The depolarizing channels are realized as more general CPTP maps, which necessarily involve a sum over more than one Kraus operator. This requires a significant extension of the discussion from the previous section. Each term in the sum contributes to the effect of noise on the density matrix. To demonstrate that a smooth map between the noisy and noise-free density matrices is possible, a homotopy for each term would be needed. However, even if such a collection of homotopies is found, this is not sufficient. The Stokes parameters, as defined in (\ref{StokesP}), are linear in the density matrix, whereas the Skyrmion number (\ref{SkNumb}) is cubic in the Stokes parameters. This non-linear dependence of the Skyrmion number on the density matrix implies that there will be interference between terms corresponding to different Kraus operators. Thus, even if a collection of homotopies exists, it does not allow us to draw conclusions about these interference effects. Consequently, we will rely on numerical methods to evaluate the effect of the quantum channel on the Skyrmion number. To provide a comprehensive treatment, we consider the complete list of channels given in the textbook \cite{nielsen2001quantum}. Our results consistently indicate that the Skyrmion number is surprisingly robust against the addition of noise.

The channels described in \cite{nielsen2001quantum} have been shown to be relevant for the description of depolarization in other studies. For example, master equations that describe quantum light depolarization have been considered in \cite{klimov2008quantum}. These models use decoherence to model the decorrelation that occurs during depolarization. While decoherence is usually accompanied by dissipation, depolarization specifically considers pure decoherence (or dephasing), where energy dissipation is negligible. In the terminology of \cite{nielsen2001quantum}, the resulting depolarizing channel is a decoherence process induced by unbiased noise generating bit-flip and phase-flip errors. For the case of light, an additional subtlety arises due to the need to account for an SU(2) invariance \cite{klimov2008quantum}. 

\subsection{Bit flip channel}

The bit flip channel is described by the operators
\begin{eqnarray}
E_0&=&\sqrt{p}\left(\begin{array}{cc} 1 & 0 \\ 0 & 1 \\ \end{array}\right)\qquad E_1\,\,=\,\, \sqrt{1-p}\left(\begin{array}{cc} 0 & 1 \\ 1 & 0 \\ \end{array}\right)
\end{eqnarray}
which obey $E_0^\dagger E_0+E_1^\dagger E_1= {\bf 1}$ with ${\bf 1}$ the $2\times 2$ identity matrix. With a probability $p$ states $|H\rangle$ and $|V\rangle$ remain what they are and with a probability $1-p$, they are swapped. To get a local model, we must allow $p$ to be a smooth function of $x$ and $y$. Notice that $0\le p\le 1$. A simple choice for $p$ is
\begin{eqnarray}
p&=&(\alpha+\beta\cos(n\phi))e^{-\gamma\rho^2}\label{simplechoices}
\end{eqnarray}
where we have again employed polar coordinates. As $\rho\to\infty$ we find that $p\to 0$. The constant parameter $\gamma$ controls how quickly $p$ approaches 0. At $p=0$ the states $H$ and $V$ are swapped so that this channel is not confined to a local region but rather continues to have an effect even at $\rho=\infty$. Acting on the Stokes parameters, this channel takes
\begin{equation}
S_x\,\,\to\,\,S_x\quad
S_y\,\,\to\,\,-S_y\quad
S_z\,\,\to\,\,-S_z
\end{equation}
at large $\rho$ where $p=0$. It is simple to see that the Skyrmion number is invariant under this transformation. The constant parameters $\alpha$ and $\beta$ are needed, with $\alpha>\beta$, to enable $p$ to oscillate but remain positive. The constant parameter $n$ controls this oscillatory behaviour as $\phi$ is varied for fixed $\rho$. We only consider integer $n$. For the above choice of $p$ we can easily evaluate the action of the channel on the density matrix. Using the noisy density matrix we have verified numerically that the Skyrmion number is robust against the polarization flip channel.

\begin{figure*}[t!]
\includegraphics[width=1.0\linewidth]{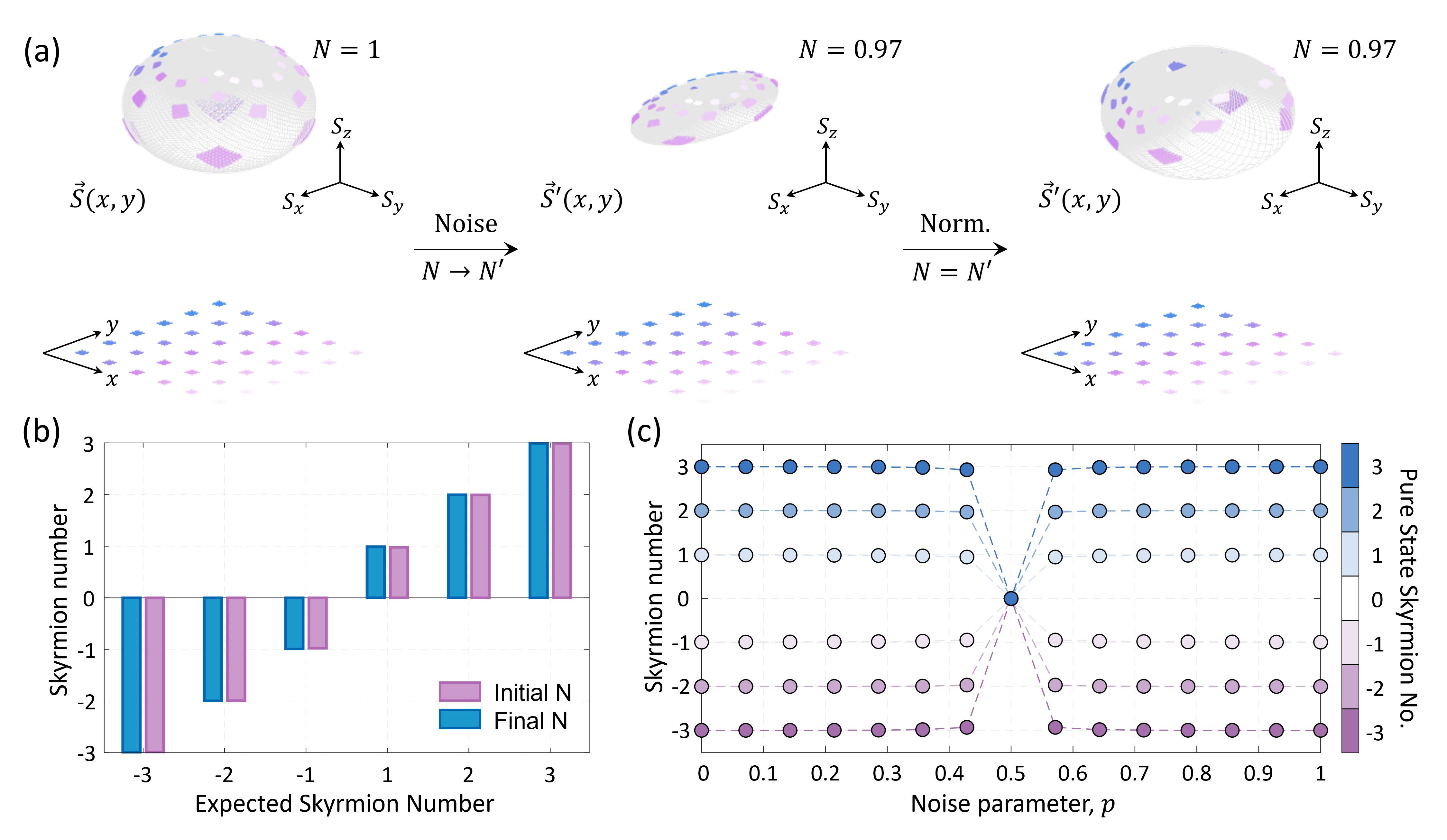}
\caption{\textbf{Skyrmion Resilience Against Constant Bit-Flip Channel.} (a) Skyrmion map for $N=1$ before (left panel) and after (middle panel) passing through a constant bit-flip channel with noise parameter $p=0.35$. The mapping is visibly distorted, $\vec{S}(x,y) \to \vec{S}'(x,y)$. To properly evaluate the Skyrmion number, $\vec{S}'(x,y)$ is normalized (right panel) such that $\vec{S}'(x,y)\cdot\vec{S}'(x,y) = 1$. The distortion due to noise is a simple shifting of points on the sphere which preserves the wrapping number, so that the calculated Skyrmion number is $N\approx0.97$. This is highlighted by matching colors to correlate regions of interest on the plane with those on the sphere, emphasizing that the movement of points on the sphere can be corrected by a coordinate transformation of the plane, i.e., $\vec{S}(x,y) \to \vec{S}'(x,y) \to \vec{S}(x',y')$. (b) Numerically calculated Skyrmion numbers before and after passing through the noisy channel ($p=0.35$) for various topologies, $N \in \{-3, -2, -1, 1, 2, 3\}$, demonstrate clear invariance of the topology to the bit-flip channel. (c) The calculated Skyrmion number plotted against the noise parameter $p$ shows the robustness of the topology across all bit-flip noise settings, with the only exception occurring at $p=0.5$, where entanglement is completely lost.} 
\label{fig:BitFlip}
\end{figure*}

There is an interesting point that deserves discussion. Taking $p$ constant, $l_1=12$ and $l_2=1$, we find the Stokes parameters, before normalization, are given by
\begin{eqnarray}
S_x&=&\frac{4 \sqrt{\frac{2}{231}} \rho ^{13} \left(1+e^{22 i \phi }\right) e^{-\frac{2 \rho ^2}{w_0^2}-11 i \phi }}{45 \pi  w_0^{15}}\cr\cr
S_y&=&-\frac{4 i \sqrt{\frac{2}{231}} (2p-1) \rho ^{13} \left(-1+e^{22 i \phi }\right) e^{-\frac{2 \rho ^2}{w_0^2}-11 i \phi }}{45 \pi  w_0^{15}}\cr\cr
S_z&=&\frac{4 (2p-1) \rho ^2 e^{-\frac{2 \rho ^2}{w_0^2}} \left(467775 w_0^{22}-2 \rho ^{22}\right)}{467775 \pi  w_0^{26}}
\end{eqnarray}
It is clear that choosing $p={1\over 2}$ we have $S_y=S_z=0$ and the Skyrmion number vanishes. Thus, for a constant $p={1\over 2}$ the Skyrmion number is completely destroyed.  The reason for this anomalous behaviour is simply that for $p={1\over 2}$ the channel sums the original density matrix with the density matrix obtained by swapping $|V\rangle$ and $|H\rangle$. This obviously renders the density matrix symmetric so that it admits a decomposition
\begin{eqnarray}
\rho &=&{1\over 2} \sigma_0 +c_1\sigma_1
\end{eqnarray}
with one undetermined coefficient $c_1$. From the formula (\ref{StokesP}) it is now apparent why both $S_y$ and $S_z$ vanish. It is also clear that this does not happen at any other value $p\ne {1\over 2}$. Pertinent to this is how likely is it that realistic conditions might result in exactly $p={1\over 2}$?  First, we can anticipate that the value of $p$ will fluctuate. Second, note that the isolated points or even lines when $p={1\over 2}$ are all lower dimensional (0-dimensional or 1-dimensional) compared to the set (2-dimensional) we are integrating over, so they are all sets of measure zero. Changing an integrand on a set of measure zero will not affect the value of an integral.  These theoretical and practical considerations lead us to surmise that this $p={1\over 2}$ anomalous behaviour will not play any role, and consequently that the Skyrmion number will remain robust in such channels.

\begin{figure}[t!]
\includegraphics[width=1.0\linewidth]{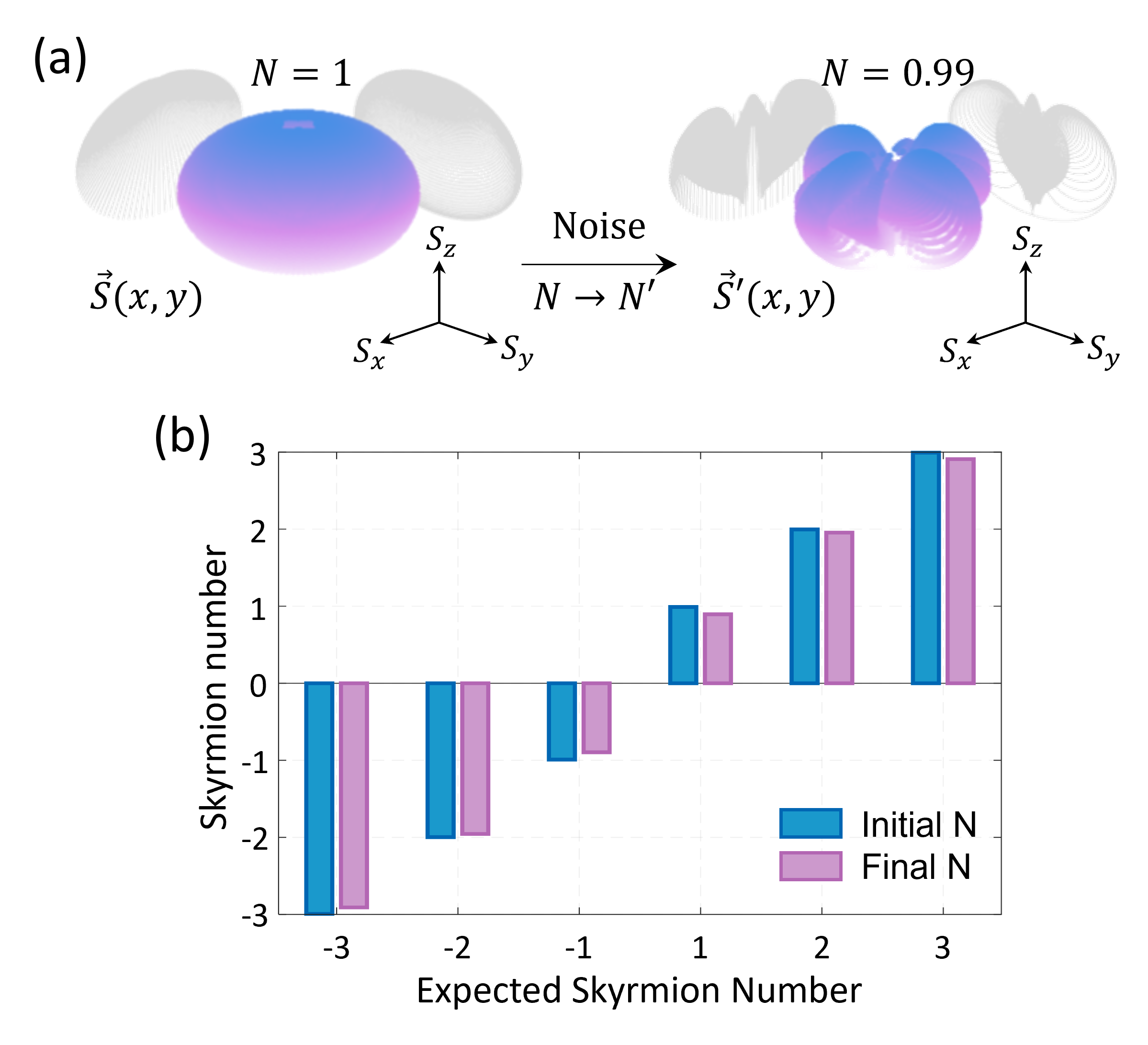}
\caption{\textbf{Skyrmion Resilience Against a Spatially Varying Bit-Flip Channel.} (a) Skyrmion map for $N=1$ before (left panel) and after (right panel) passing through a spatially varying bit-flip channel. Although the mapping is significantly distorted, $\vec{S}(x,y) \to \vec{S}'(x,y)$, the calculated Skyrmion number remains close to the expected value at $N\approx0.99$, indicating that the topology is preserved despite the deformation. (b) Numerically calculated Skyrmion numbers for various topologies, $N \in \{-3, -2, -1, 1, 2, 3\}$, before and after passing through the spatially varying bit-flip channel, demonstrating the clear invariance of the topology even under spatially dependent noise.}
\label{fig:BitFlipSpatial}
\end{figure}

Further numerical results for various Skyrmion topologies passed through both constant and spatially varying bit-flip channels are presented in Fig.\ref{fig:BitFlip} and Fig.\ref{fig:BitFlipSpatial}, respectively. In Fig.~\ref{fig:BitFlip} (a), the initial map $\vec{S}(x,y)$ with a Skyrmion number $N=1$ undergoes transformation to $\vec{S}'(x,y)$ after passing through a bit-flip channel with a noise probability of $p=0.35$. This introduces noticeable distortion in the spherical geometry of $\vec{S}(x,y)$, as the $S_y$ and $S_z$ components are scaled by a factor of $1-2p$, giving the geometry an elliptical appearance. However, following a renormalization to ensure $\vec{S}'(x,y)\cdot\vec{S}'(x,y) = 1$, we observe that the noise-induced transformation simply rotates the Stokes vectors, shifting the regions of interest (ROIs) on the sphere without altering the overall coverage. Consequently, the Skyrmion number remains unchanged at $N\approx0.97$. Figure \ref{fig:BitFlip} (b) shows the numerical results for initial and final states possessing Skyrmion numbers $N \in \{-3,-2,-1,1,2,3\}$ after passing through the same constant bit-flip channel, while Fig.\ref{fig:BitFlip} (c) explores these results across various bit-flip noise settings. The data reveal that the topology remains fully invariant to bit-flip noise, with the Skyrmion number only vanishing at the singular limit of $p=0.5$. At this limit, the state becomes unentangled, and by definition, $N=0$. Intuitively, we find that at $p=0.5$, $S_y = S_z = 0$, meaning that after renormalization, $\vec{S} = [1, 0, 0]^T$ for all positions $(x, y)$, so that the initial topology has been lost. In Fig.\ref{fig:BitFlipSpatial} (a), numerical results illustrate the distortion of $\vec{S}(x,y)$ with $N=1$ as it passes through a spatially varying bit-flip channel with a spatially dependent noise parameter, $p(x,y)$, defined by (\ref{simplechoices}). Although the new map $\vec{S}'(x,y)$ parametrizes a highly distorted geometry relative to the original spherical form, the initial Skyrmion number of $N\approx0.99$ is preserved, as expected, due to the smooth spatial dependence of the noise. Figure \ref{fig:BitFlipSpatial} (b) confirms this robustness across different topological numbers passed through the same noise channel, demonstrating that Skyrmion topology remains invariant under both varying noise levels and spatial dependent noise.

\subsection{Phase flip channel}

The phase flip channel is described by the operators
\begin{eqnarray}
E_0&=&\sqrt{p}\left(\begin{array}{cc} 1 & 0 \\ 0 & 1 \\ \end{array}\right)\qquad E_1\,\,=\,\, \sqrt{1-p}\left(\begin{array}{cc} 1 & 0 \\ 0 & -1 \\ \end{array}\right)
\qquad\end{eqnarray}
which obey $E_0^\dagger E_0+E_1^\dagger E_1={\bf 1}$ with ${\bf 1}$ the $2\times 2$ identity matrix and again we must allow $p$ to be a smooth function of $x$ and $y$. It is again useful to consider the simple choice (\ref{simplechoices}). With this choice our numerical results confirm that the Skyrmion number is robust against noise introduced by the phase flip channel.

In the previous subsection, we found that the bit flip channel with the specific parameter choice $p={1\over 2}$ completely destroyed the Skyrmion number, reducing it to zero. There is a similar result for the phase flip channel: choosing $p={1\over 2}$ we have $S_x=S_y=0$ and the Skyrmion number vanishes. Just as for the bit flip channel, even though the parametrization (\ref{simplechoices}) leads to $p={1\over 2}$ at some isolated points or even lines, these bad values do not disturb the Skyrmion number under realistic situations: once again, in practice the value of $p$ will fluctuate so we do not expect that this $p={1\over 2}$ anomalous behaviour will play any role.
\begin{figure*}[t!]
\includegraphics[width=1.0\linewidth]{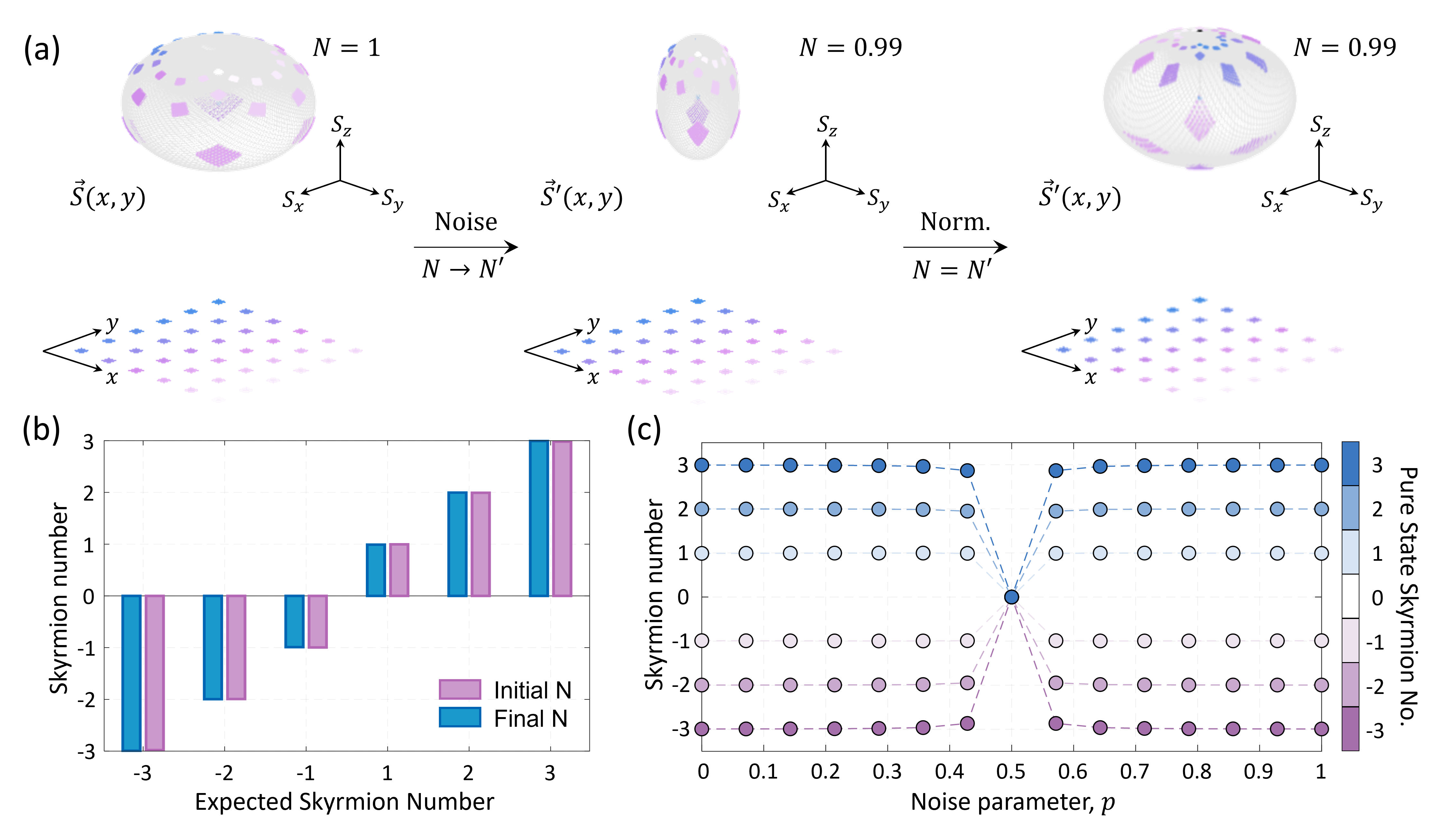}
\caption{\textbf{Skyrmion Resilience Against Constant Phase Flip Channels.} (a) Skyrmion map for $N=1$ before (left panel) and after (middle panel) passing through a constant phase flip channel with a noise parameter of $p=0.35$. The mapping is clearly distorted, $\vec{S}(x,y) \to \vec{S}'(x,y)$. However, after normalizing $\vec{S}'(x,y)$ such that $\vec{S}'(x,y) \cdot \vec{S}'(x,y) = 1$ (right panel), it is evident that the points have merely shifted without altering the wrapping number. The calculated Skyrmion number remains close to the original, $N=0.99$. This is further illustrated by matching colored regions of interest (ROIs) on the plane with corresponding regions on the sphere, highlighting the shifts that can be corrected by coordinate transformation, i.e., $\vec{S}(x,y) \to \vec{S}'(x,y) \to \vec{S}(x',y')$. (b) Numerically calculated Skyrmion numbers (before and after passing through a noisy channel with $p=0.35$) compared to the expected Skyrmion numbers for various topologies, $N \in \{-3, -2, -1, 1, 2, 3\}$, showing clear invariance to the phase flip channel. (c) Skyrmion number plotted against the noise parameter $p$, demonstrating topological invariance across all phase flip noise settings except at $p=0.5$, where entanglement is completely lost, reducing $N$ to 0.}
\label{fig:PhaseFlip}
\end{figure*}

\begin{figure}[t!]
\includegraphics[width=1.0\linewidth]{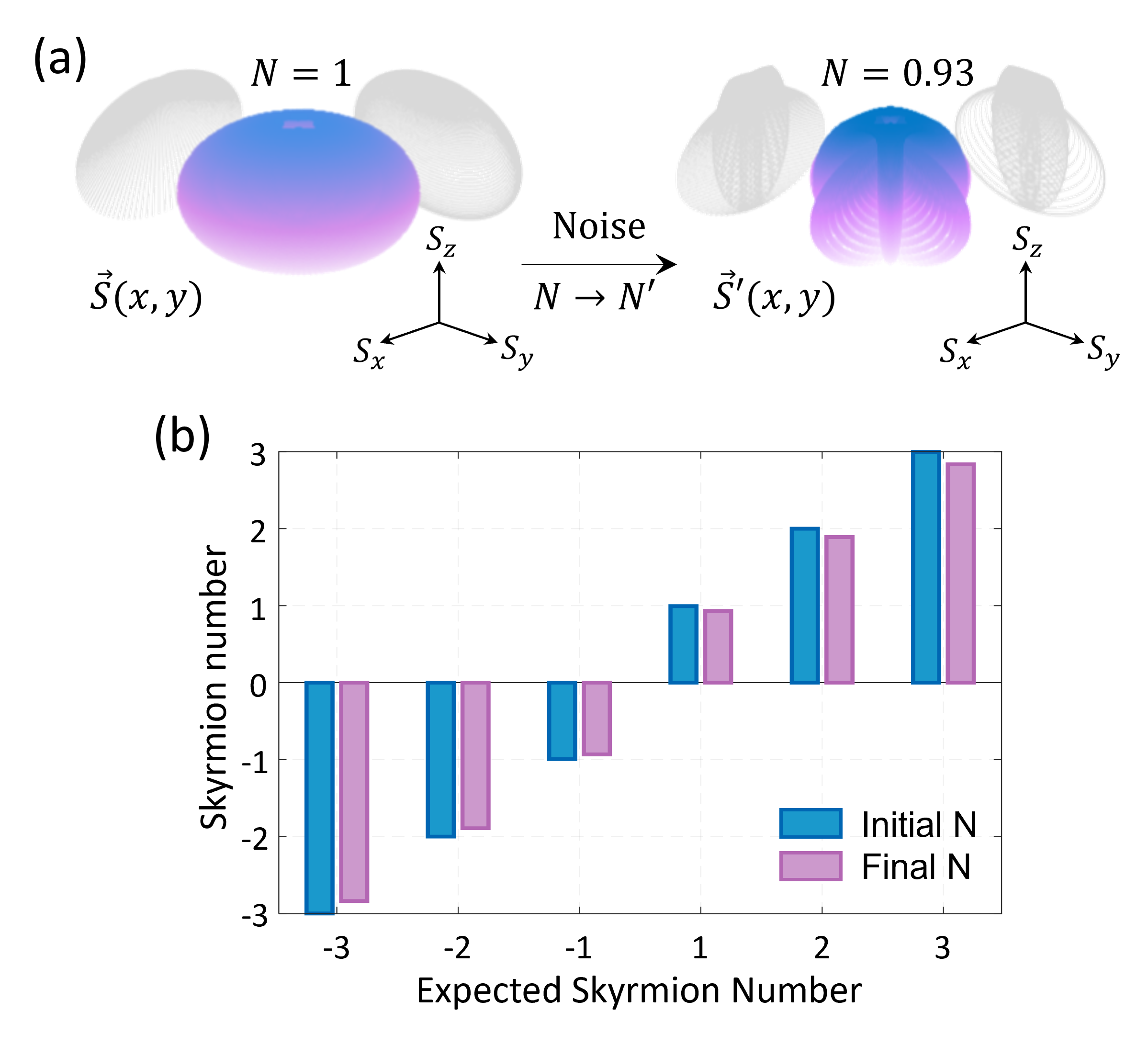}
\caption{\textbf{Skyrmion Resilience Against a Spatially Varying Phase Flip Channel.} (a) Skyrmion map for $N=1$ before (left panel) and after (right panel) passing through a spatially varying phase flip channel. Although the mapping is significantly distorted, $\vec{S}(x,y) \to \vec{S}'(x,y)$, the calculated Skyrmion number remains close to the expected value, with $N\approx0.93$. This suggests that despite the heavy distortion, the topological structure is largely preserved. (b) Numerically calculated Skyrmion numbers before and after passing through the noisy channel are compared to the expected values for various topologies, $N \in \{-3,-2,-1,1,2,3\}$. The results demonstrate a clear invariance of the topological number, confirming the robustness of the Skyrmion topology to spatially varying phase flip noise.}
\label{fig:PhaseFlipSpatial}
\end{figure}

Further numerical results for various Skyrmion topologies passed through constant and spatially varying phase-flip channels are shown in Fig.\ref{fig:PhaseFlip} and Fig.\ref{fig:PhaseFlipSpatial}, respectively. In Fig.\ref{fig:PhaseFlip} (a), the initial map $\vec{S}(x,y)$, with Skyrmion number $N=1$, is passed through a phase-flip channel with noise parameter $p=0.35$ to give the noisy $\vec{S}'(x,y)$ signal. This channel induces a noticeable distortion in the spherical geometry of the initial wave function. This is as expected as the $S_x$ and $S_y$ components are scaled by a constant factor of $1-2p$, giving the appearance of an elliptical deformation. However, after renormalizing the Stokes parameters so that $\vec{S}'(x,y)\cdot\vec{S}'(x,y)=1$, it is evident that the noise simply rotates the Stokes vectors. Consequently, the coverage of the sphere is intact and the Skyrmion number is preserved. In Fig.\ref{fig:PhaseFlip} (b), numerical results for Skyrmion numbers $N \in \{-3,-2,-1,1,2,3\}$, before and after passing through the constant phase-flip channel, show that the topology is unchanged. Fig.~\ref{fig:PhaseFlip} (c) illustrates this invariance across a range of phase-flip noise settings. The topology is unaffected by phase-flip noise, except at the critical point $p=0.5$, where the entanglement vanishes, and by definition, $N=0$. Intuitively, at this point, $S_x=S_y=0$, and after renormalization, $\vec{S}=[0,0,1]^T$ for all positions $(x,y)$, indicating a complete collapse of the original topology. In Fig.\ref{fig:PhaseFlipSpatial} (a), numerical results illustrate the deformation of $\vec{S}(x,y)$ with $N=1$ after passing through a spatially varying phase-flip channel, where the noise parameter $p(x,y)$ is defined according to (\ref{simplechoices}). Despite the apparent significant distortion of the geometry of the wave function when compared to the initial state, the Skyrmion number is intact at $N\approx0.99$, as expected for a noise parameter that varies smoothly with changing values of the spatial coordinates. Figure \ref{fig:PhaseFlipSpatial} (b) presents results for different topological numbers passed through the same spatially varying noise channel, confirming complete robustness of the Skyrmion topology to phase-flip noise across different noise levels and spatial variations.

\subsection{Depolarizing channel}

The depolarizing channel is described by the quantum operation ($0\le p\le 1$)
\begin{eqnarray}
{\cal E}(\rho)&=&{p\over 2}{\bf 1}+(1-p)\rho\label{DepChan}
\end{eqnarray}
This channel is very similar to the case of isotropic noise considered in \cite{ornelas2024topological}. The present analysis generalizes the discussion of \cite{ornelas2024topological} since we allow $p$ to be a smooth function of $x$ and $y$. Since our density matrix lives in the tensor product of a two dimensional polarization Hilbert space, with a two dimensional position Hilbert space, we must be careful to choose the coefficient of the term ${\bf 1}$ in such a way that our state remains normalizable. A suitable choice is to take $p$ to be of the form
\begin{eqnarray}
p(\rho,\phi)&=&(\beta+\gamma \cos (n\phi )) \exp \left(-\frac{\rho ^2}{u_0^2}\right)\label{eq:DepSimpleEq}
\end{eqnarray}
with $\beta$ and $\gamma$ chosen to ensure that $0<p\le 1$. With the above choice we find that $p\to 0$ as $\rho\to\infty$ so that the channel acts in a local region centred on $\rho=0$. The parameter $n$ controls how $p$ fluctuates at fixed $\rho$ as $\phi$ is varied. We assume that $n$ is integer. Our numerical results confirm that the Skyrmion number is completely robust to the depolarizing channel. The discussion is exactly as in \cite{ornelas2024topological}. That analysis proves that the depolarizing channel induces a smooth deformation of the Skyrmion state that leaves the topology unchanged. The topology only collapses when the state is maximally mixed and no entanglement remains.

In the depolarizing channel the density matrix is perturbed by a maximally mixed state which corresponds to adding a term proportional to the identity. This maximally mixed term gives a vanishing contribution to the Stokes parameters, so that in the end the net effect is to multiply the Stokes parameters by an overall constant. This change in the magnitude of the Stokes vector is a smooth deformation of the Stokes parameters, so that for this channel it is also possible to give a homotopy between the noisy and noise free Stokes parameters. This is the only depolarizing channel we have considered that admits a homotopy analysis. The reason why the homotopy analysis continues to be useful is simply that the maximally mixed term (i.e. the term on the RHS of (\ref{DepChan}) proportional to ${\bf 1}$) gives a vanishing contribution to the Stokes parameters so there are no interference effects between the signal and noise in the non-linear formula (\ref{SkNumb}).

\begin{figure*}[t!]
\includegraphics[width=1.0\linewidth]{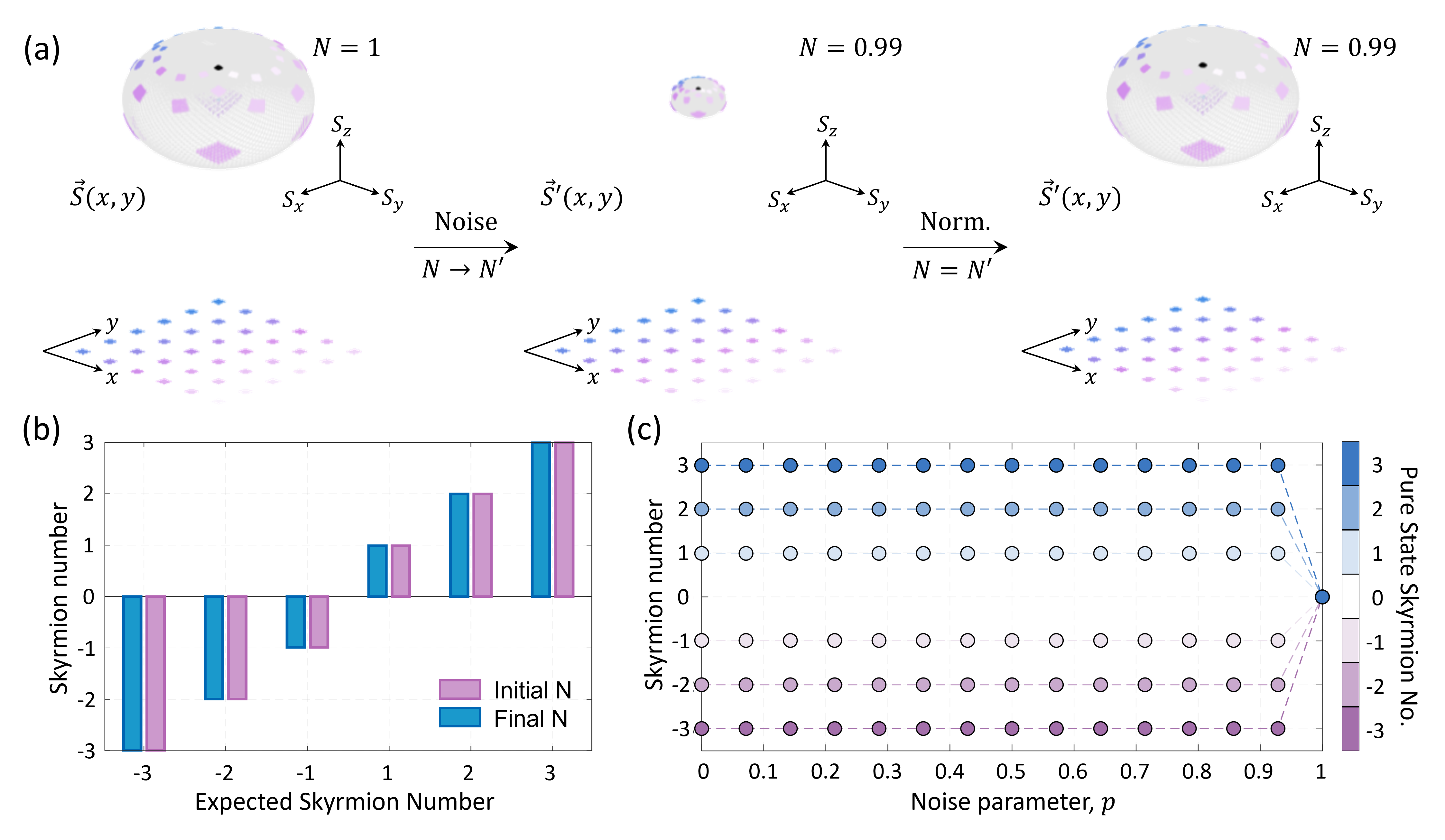}
\caption{\textbf{Skyrmion resilience against constant depolarizing channel.} (a) Skyrmion map defined by the density matrix, for $N=1$ before (left panel) and after (middle panel) passing through a constant depolarizing channel with a noise parameter $p=0.35$. The mapping $\vec{S}(x,y) \to \vec{S}'(x,y)$ shows that the density matrix has suffered a clear distortion due to depolarizing noise, which reduces the magnitude of $\vec{S}(x,y)$, i.e., $\vec{S}(x,y)\cdot\vec{S}(x,y)<1$. After renormalizing $\vec{S}'(x,y)$ such that $\vec{S}'(x,y)\cdot\vec{S}'(x,y)=1$ (right panel), the Stokes vectors recover their spherical orientation, and the Skyrmion number remains unchanged at $N\approx0.99$. (b) Numerically calculated Skyrmion numbers (before and after passing the wave function through the noisy channel with $p=0.35$) compared to expected Skyrmion numbers for various topologies, $N \in \{-3,-2,-1,1,2,3\}$, demonstrate the topology's invariance to depolarizing noise. (c) The calculated Skyrmion number as a function of the noise parameter $p$ shows the robustness of the topology across all depolarizing noise levels, except at $p=1$, where the state becomes completely mixed, and entanglement is lost which reduces the Skyrmion number to $N=0$.}
\label{fig:Depolarizing}
\end{figure*}

\begin{figure}[t!]
\includegraphics[width=1.0\linewidth]{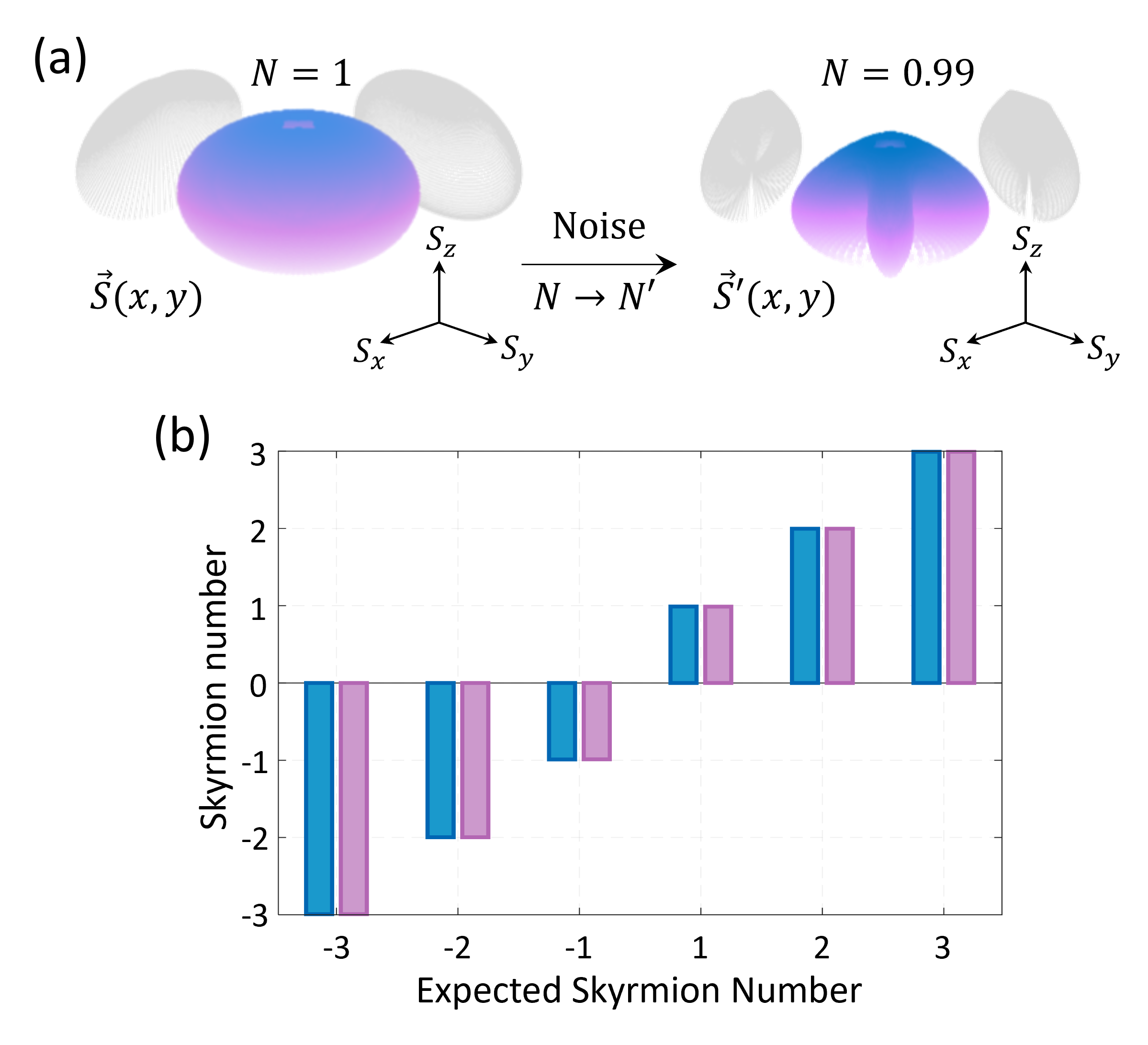}
\caption{{\bf Skyrmion resilience against a spatially varying depolarizing channel.} (a) Skyrmion map for $N=1$ before (left panel) and after (right panel) passing the density matrix through a spatially varying depolarizing channel. Despite significant distortion in the mapping, $\vec{S}(x,y) \to \vec{S}'(x,y)$, the calculated Skyrmion number remains robust, with a value of $N\approx0.99$, close to the expected value. (b) Numerically calculated Skyrmion number, both before and after the state passes through the noisy channel, plotted against the expected Skyrmion number for various topologies, $N \in \{-3,-2,-1,1,2,3\}$. The results demonstrate a clear invariance of topology, even under the influence of the spatially varying depolarizing channel.}
\label{fig:DepolarizingSpatial}
\end{figure}

Further numerical results for various Skyrmion topologies passed through constant and spatially varying depolarizing channels are shown in Fig.\ref{fig:Depolarizing} and Fig.\ref{fig:DepolarizingSpatial}, respectively. In Fig.~\ref{fig:Depolarizing} (a), application of the depolarizing channel with parameter $p=035$ to the initial state $\vec{S}(x,y)$, with a Skyrmion number $N=1$, produces the noisy state $\vec{S}'(x,y)$. This modulation reduces $\vec{S}'(x,y)\cdot\vec{S}'(x,y)<1$, as expected from the uniform scaling of all components by the factor $1-p$. This results in a reduction of the radius of the original spherical geometry encoded in the Skyrmion wave function. After renormalizing the Stokes parameters so that $\vec{S}'(x,y)\cdot\vec{S}'(x,y)=1$, it is evident that the depolarizing noise does not affect the orientation of the Stokes vectors. Consequently, the Skyrmion number is unaltered, $N\approx0.99$. In Fig.\ref{fig:Depolarizing} (b), numerical results are presented for Skyrmion numbers $N \in \{-3,-2,-1,1,2,3\}$ before (initial) and after (final) passing the state through the constant depolarizing channel, demonstrating the invariance of the topology. Figure \ref{fig:Depolarizing} (c) shows how the Skyrmion number remains robust across various depolarizing noise settings, with the topology only breaking down at the extreme limit of $p=1$. At this point, the state is completely mixed and entanglement is lost resulting in $N=0$. In this scenario, $S_x=S_y=S_z=0$ and the initial topology can not be recovered. In Fig.\ref{fig:DepolarizingSpatial} (a), numerical results illustrate the distortion of $\vec{S}(x,y)$, with a Skyrmion number $N=1$, after passing through a spatially varying depolarizing channel, where the noise parameter $p(x,y)$ is a smooth function of spatial coordinates defined in (\ref{eq:DepSimpleEq}). Despite the apparent significant distortion of the geometry encoded in the noisy wave function, as compared to the original spherical structure, the Skyrmion number remains intact at $N\approx0.99$. This is again not surprising as the noise varies smoothly. Figure \ref{fig:DepolarizingSpatial} (b) presents results for different topological numbers passed through the same spatially varying noise channel, confirming the robustness of Skyrmion topologies against depolarizing noise at various noise levels and with spatially varying noise.

\subsection{Amplitude damping channel}

The amplitude damping channel is described by the operators
\begin{eqnarray}
E_0&=&\left(\begin{array}{cc} 1 & 0 \\ 0 & \sqrt{1-p} \\ \end{array}\right)\qquad E_1\,\,=\,\, \left(\begin{array}{cc} 0 & \sqrt{p} \\ 0 & 0 \\ \end{array}\right)
\end{eqnarray}
which obey $E_0^\dagger E_0+E_1^\dagger E_1= {\bf 1}$ with ${\bf 1}$ the $2\times 2$ identity matrix and again we must allow $p$ to be a smooth function of $x$ and $y$. Using this channel, the noisy density matrix is given by
\begin{eqnarray}
\rho&=&\left(\begin{array}{cc} \rho_{11}+p\rho_{22} & \sqrt{1-p}\rho_{12} \\ \sqrt{1-p}\rho_{21} & (1-p)\rho_{22} \\ \end{array}\right)
\end{eqnarray}
The form of the channel parameter $p$ assumed in the numerical analysis is
\begin{eqnarray}
p&=&\exp \left(-\frac{\rho ^2}{u_0^2}\right) (\alpha \cos (n\phi )+\beta)
\end{eqnarray}
When $p=0$ this channel acts as the identity. Notice that as $\rho\to\infty$, $p\to 0$ with the fall off controlled by the parameter $u_0$. We choose $\beta>\alpha$ to ensure that $p$ remains positive. The integer $n$ controls how $p$ oscillates for fixed $\rho$ as $\phi$ is varied. Our numerical results confirm that the Skyrmion number is completely robust to amplitude damping.

For the amplitude damping channel, the Stokes parameters with a constant $p$, $l_1=12$ and $l_2=1$ are given by
\begin{eqnarray}
S_x&=&\frac{4 \sqrt{\frac{2}{231}} \sqrt{1-p} \rho ^{13} \left(1+e^{22 i \phi }\right) e^{-\frac{2 \rho ^2}{w_0^2}-11 i \phi }}{45 \pi  w_0^{15}}\cr\cr
S_y&=&-\frac{4 i \sqrt{\frac{2}{231}} \sqrt{1-p} \rho ^{13} \left(-1+e^{22 i \phi }\right) e^{-\frac{2 \rho ^2}{w_0^2}-11 i \phi }}{45 \pi  w_0^{15}}\cr\cr
S_z&=&\frac{4 \rho ^2 e^{-\frac{2 \rho ^2}{w_0^2}} \left(2 (2p-1) \rho ^{22}+467775 w_0^{22}\right)}{467775 \pi  w_0^{26}}
\end{eqnarray}
Notice that at $p={1\over 2}$ we have $S_z=0$ and at $p=1$ both $S_x=0$ and $S_y=0$. The Skyrmion number vanishes for both of these values. The discussion is again parallel to the bit flip and phase flip channels: even though the parameterization we use for $p$ does pass through $p={1\over 2}$ and $p=1$ values, the Skyrmion number is robust to the amplitude damping channel and we do not expect these particular channel parameter values to play any role in practise.

\begin{figure*}[t!]
\includegraphics[width=1.0\linewidth]{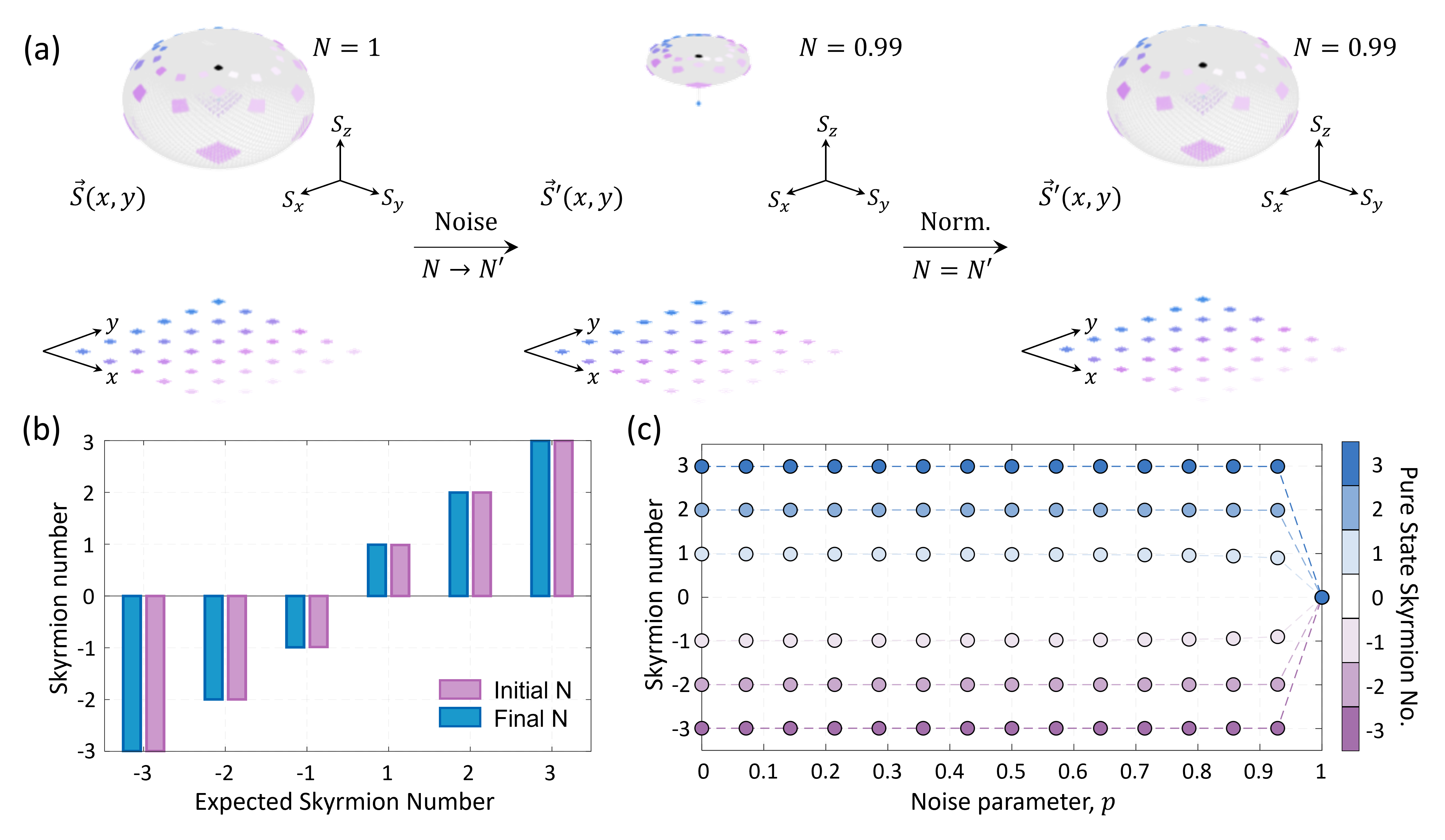}
\caption{{\bf Skyrmion resilience against constant amplitude damping channel.} (a) Skyrmion map for $N=1$ before (left panel) and after (middle panel) passing the desnity matrix through a constant amplitude damping channel with a noise parameter of $p=0.35$. The mapping $\vec{S}(x,y) \to \vec{S}'(x,y)$ shows clear distortion. To properly evaluate the Skyrmion number, the map is renormalized (right panel) so that $\vec{S}'(x,y) \cdot \vec{S}'(x,y) = 1$. Despite the distortion, the wrapping number is preserved, with a calculated Skyrmion number of $N\approx0.99$. Matching colors highlight correlated regions of interest on the plane and sphere, exhibiting the movement of points on the sphere. This can be corrected for with a coordinate transfdormation as $\vec{S}(x,y) \to \vec{S}'(x,y) \to \vec{S}(x',y')$. (b) Numerically calculated Skyrmion numbers before and after passing the desnity matrix through the noisy channel with $p=0.35$, plotted against the expected Skyrmion number for various topologies, $N \in \{-3,-2,-1,1,2,3\}$. This demonstrates that the Skyrmion number is robust to noise from the amplitude damping channel. (c) The Skyrmion number plotted against the noise parameter, $p$, demonstrates topological invariance across all noise settings except at $p=1$, where the state entanglement is completely lost and the Skyrmion number vanishes.}
\label{fig:AmpDamp}
\end{figure*}

\begin{figure}[t!]
\includegraphics[width=1.0\linewidth]{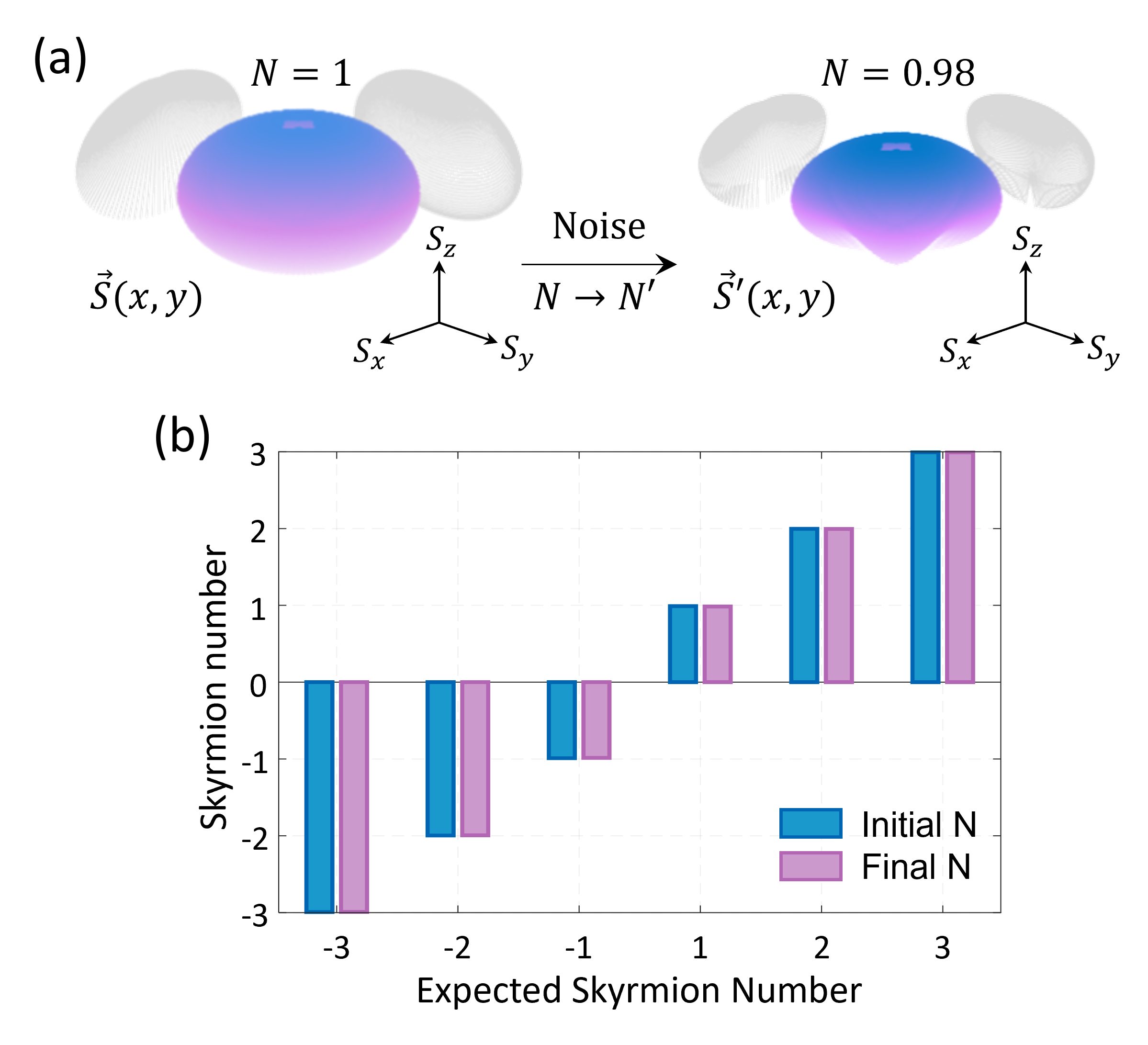}
\caption{{\bf Skyrmion resilience against a spatially varying amplitude damping channel.} (a) Skyrmion map for $N=1$ before (left panel) and after (right panel) passing the density matrix through a spatially varying amplitude damping channel. Despite significant distortion in the mapping, $\vec{S}(x,y) \to \vec{S}'(x,y)$, the calculated Skyrmion number remains essentially unchanged at $N\approx0.98$. (b) Numerically calculated Skyrmion numbers before and after passing the density matrix through the noisy channel are compared against expected values for various Skyrmion topologies, $N \in \{-3,-2,-1,1,2,3\}$. The results demonstrate clear topological invariance of the Skyrmion number under the spatially varying amplitude damping channel.}
\label{fig:AmpDampSpatial}
\end{figure}

Further numerical results for various skyrmion topologies resulting after the state is passed through both constant and spatially varying amplitude damping channels are shown in Fig.\ref{fig:AmpDamp} and Fig.\ref{fig:AmpDampSpatial}, respectively. In Fig.~\ref{fig:AmpDamp} (a), the initial map $\vec{S}(x,y)$ with Skyrmion number $N=1$ is transformed into $\vec{S}'(x,y)$ after passing the density matrix through an amplitude damping channel with $p=0.35$. This produces a geometric distortion of the sphere encoded by the state, with points shifted toward the north pole while maintaining $\vec{S}'(x,y)\cdot\vec{S}'(x,y)\leq 1$. This behavior is expected since the components of the map transform as follows: $S_x' = S_x\sqrt{1-p}$, $S_y' = S_y\sqrt{1-p}$, and $S_z' = p + S_z(1-p)$, indicating that the $x,y$ components are scaled and $S_z$ is shifted by a positive constant. At $p=1$, all points collapse to the north pole, i.e., $\vec{S}'(x,y) = [0, 0, 1]^T$. After renormalizing the Stokes parameters it is evident that noise does not affect the Stokes vectors' orientation, so the Skyrmion number is unchanged at $N\approx0.99$. In Fig.\ref{fig:AmpDamp} (b), the numerical results show Skyrmion numbers for various topologies $N \in \{-3,-2,-1,1,2,3\}$ before and after passing through the constant amplitude damping channel, establishing the topological robustness. Fig.\ref{fig:AmpDamp} (c) displays Skyrmion numbers across different amplitude damping noise levels, revealing that the topology is unaffected by noise except at the extreme limit where $p=1$. At this point, the state is completely mixed and loses all entanglement which reduces the Skyrmion number to $N=0$. Intuitively, at $p=1$, $S_x=S_y=S_z=0$ and the initial topology can not be recovered. In Fig.\ref{fig:AmpDampSpatial} (a), numerical results are shown for the distortion of $\vec{S}(x,y)$ with Skyrmion number $N=1$ induced by passing the density matrix through a spatially varying amplitude damping channel with a spatially dependent noise parameter $p(x,y)$, defined according to (\ref{eq:DepSimpleEq}). Despite significant geometric distortion compared to the initial spherical geometry, the skyrmion number remains $N\approx0.99$. This is a consequence of the fact that the noise parameter is a smooth function of the spatial coordinates. In Fig.\ref{fig:AmpDampSpatial} (b), results for various topologies are shown, demonstrating complete robustness of the skyrmion topology to amplitude damping noise at different noise levels and with spatial variation. These results confirm the invariance of skyrmion topology against amplitude damping noise across a range of conditions.

\subsection{Phase damping channel}

The phase damping channel is described by the operators
\begin{eqnarray}
E_0&=&\left(\begin{array}{cc} 1 & 0 \\ 0 & \sqrt{1-p} \\ \end{array}\right)\qquad E_1\,\,=\,\, \left(\begin{array}{cc} 0 & 0\\ 0 & \sqrt{p} \\ \end{array}\right)
\end{eqnarray}
which obey $E_0^\dagger E_0+E_1^\dagger E_1= {\bf 1}$ with ${\bf 1}$ the $2\times 2$ identity matrix and again we must allow $p$ to be a smooth function of $x$ and $y$. The form of the channel parameter assumed for the numerical analysis is
\begin{eqnarray}
p&=&\beta\exp \left(-\frac{\rho ^2}{u_0^2}\right) (\nu \cos (n\phi )+\gamma)
\end{eqnarray}
and we choose parameters so that $p\le1$. Notice that when $p=0$ this channel acts as the identity. This channel again only acts non-trivially in a local region centred on $\rho=0$, with the fall off of $p$ controlled by $u_0$. Our numerical results demonstrate that the Skyrmion number is completely robust to the phase damping channel. Computing the Skyrmion number for $p$ constant, with $l_1=12$ and $l_2=1$ we find the Stokes parameters
\begin{eqnarray}
S_x&=&\frac{4 \sqrt{\frac{2}{231}} \sqrt{1-p} \,\rho ^{13} \left(1+e^{22 i \phi }\right) e^{-\frac{2 \rho ^2}{w_0^2}-11 i \phi }}{45 \pi  w_0^{15}}\cr\cr
S_y&=&-\frac{4 i \sqrt{\frac{2}{231}} \sqrt{1-p}\, \rho ^{13} \left(-1+e^{22 i \phi }\right) e^{-\frac{2 \rho ^2}{w_0^2}-11 i \phi }}{45 \pi  w_0^{15}}\cr\cr
S_z&=&\frac{4 \rho ^2 e^{-\frac{2 \rho ^2}{w_0^2}} \left(467775 w_0^{22}-2 \rho ^{22}\right)}{467775 \pi  w_0^{26}}
\end{eqnarray}
For $p=1$ $S_x$ and $S_y$ vanish, so that the Skyrmion number is zero. Once again we do not expect this particular channel parameter values to play any role in practise.

\begin{figure*}[t!]
\includegraphics[width=1.0\linewidth]{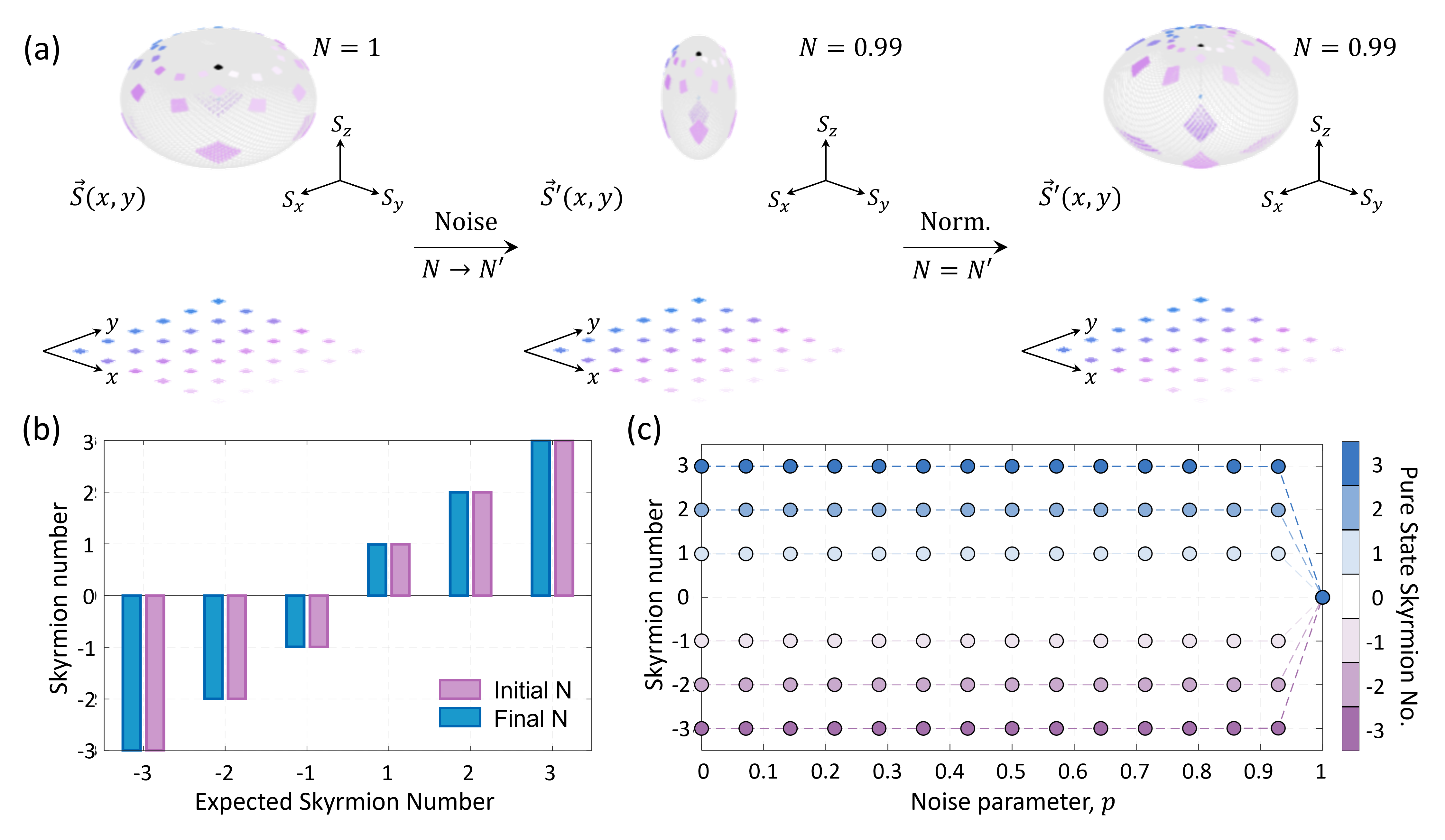}
\caption{\textbf{Skyrmion resilience against constant phase damping channel.} (a) Skyrmion map for $N=1$ before (left panel) and after (middle panel) passing the density matrix through a constant phase damping channel with noise parameter $p=0.35$. The mapping is visibly distorted, with $\vec{S}(x,y) \to \vec{S}'(x,y)$. Despite the distortion, the Skyrmion number when evaluated for the normalized but distorted map, remains nearly unchanged, $N = 0.99$. This is explained by noting that points on the sphere have merely shifted, as can be seen from the correlated regions of interest (highlighted in matching colors). These shifts on the sphere can be corrected by a coordinate transformation acting on the plane, i.e., $\vec{S}(x,y) \to \vec{S}'(x,y) \to \vec{S}(x',y')$. (b) Numerically calculated Skyrmion numbers before and after application of the noisy channel (with $p = 0.35$) compared to the expected Skyrmion numbers for various topologies $N \in \{-3, -2, -1, 1, 2, 3\}$. This establishes the invariance of the Skyrmion number to the effects of phase damping noise. (c) The Skyrmion number is plotted against the noise parameter $p$, demonstrating the robustness of the Skyrmion topology across all phase damping noise parameter values, except at $p=1$ where entanglement is completely lost.}
\label{fig:PhaseDamp}
\end{figure*}

\begin{figure}[t!]
\includegraphics[width=1.0\linewidth]{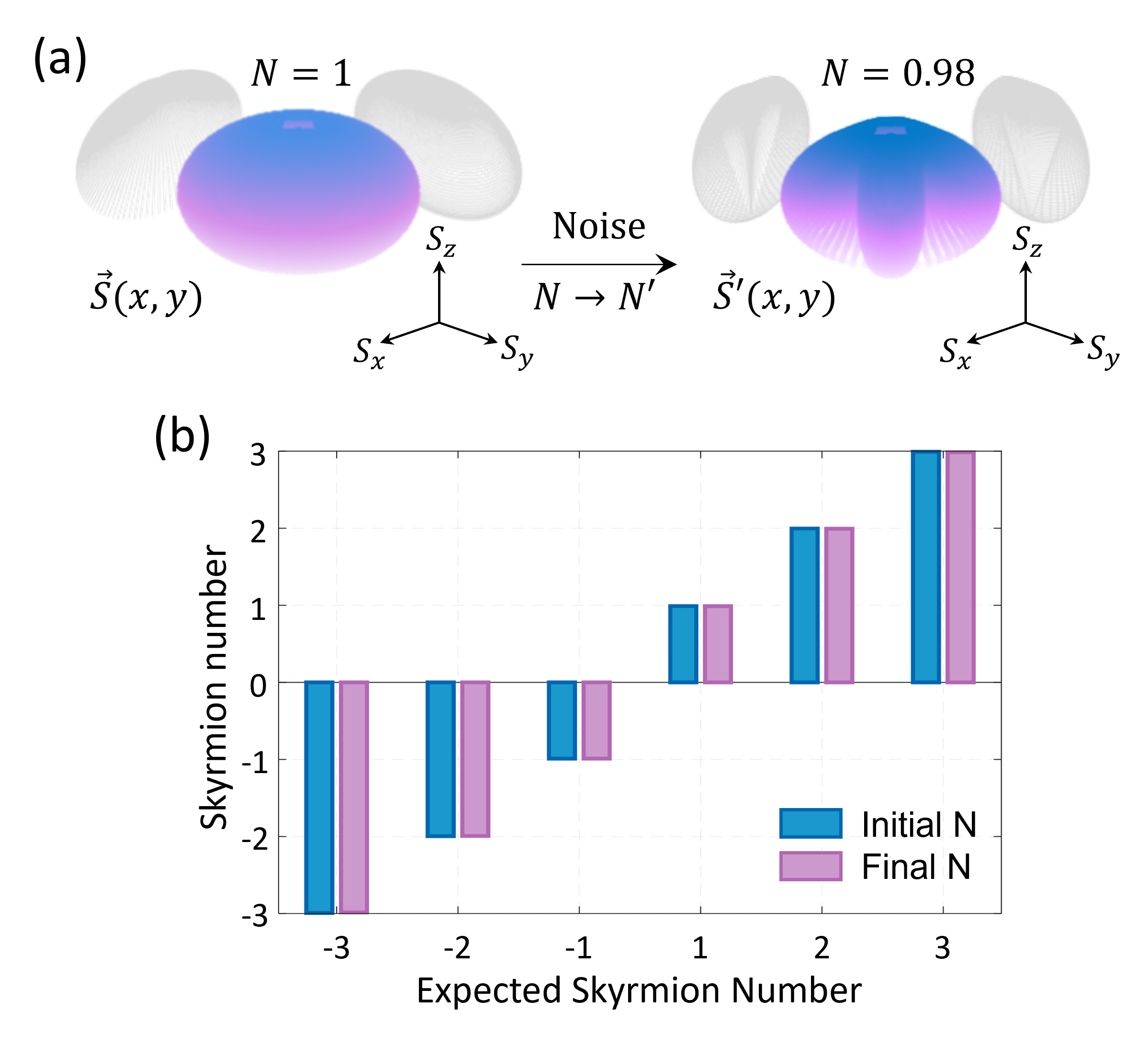}
\caption{\textbf{Skyrmion resilience against a spatially varying phase damping channel.} (a) Skyrmion map for $N=1$ before (left panel) and after (right panel) the effects of noise due to a spatially varying phase damping channel are included. Despite the significant distortion of the map, $\vec{S}(x,y) \to \vec{S}'(x,y)$, the calculated Skyrmion number remains close to the expected value at $N\approx0.98$. (b) Numerically calculated Skyrmion numbers, before and after the action of the noisy channel, are compared for various topologies, $N \in \{-3, -2, -1, 1, 2, 3\}$. The results demonstrate a clear invariance of the Skyrmion topology to the spatially varying phase damping channel.}
\label{fig:PhaseDampSpatial}
\end{figure}

Numerical results for various Skyrmion topologies computed after the density matrix is passed through constant and spatially varying phase damping channels are presented in Fig.\ref{fig:PhaseDamp} and Fig.\ref{fig:PhaseDampSpatial}, respectively. In Fig.~\ref{fig:PhaseDamp}(a), the initial Skyrmion map $\vec{S}(x,y)$ with Skyrmion number $N=1$ is transformed into $\vec{S}'(x,y)$ by a phase damping channel with noise parameter $p=0.35$. There is a noticeable distortion of the spherical geometry described by $\vec{S}(x,y)$. This is expected since the $S_x$ and $S_y$ components are scaled by a factor $\sqrt{1-p}$, much like what occured in the phase flip channel. After renormalizing the Stokes parameters it is evident that the noise only produces a rotation of the Stokes vectors, as illustrated by the shift in the positions of regions of interest (ROIs) on the sphere. Therefore, the transformation does not alter the Skyrmion number which is $N\approx0.99$. In Fig.\ref{fig:PhaseDamp}(b), numerical results for states with Skyrmion numbers $N \in \{-3,-2,-1,1,2,3\}$ are shown before and after the action of the same phase damping channel. In Fig.\ref{fig:PhaseDamp}(c), results for these states across different phase damping noise levels are displayed. These results confirm that the Skyrmion topology is invariant under phase damping noise, only collapsing at the singular limit $p=1$, where entanglement is lost and the Skyrmion number vanishes. At $p=1$, $S_x=S_y=0$ so it is not possible to recover the initial topology. Fig.\ref{fig:PhaseDampSpatial}(a) shows numerical results for the distortion of $\vec{S}(x,y)$ with Skyrmion number $N=1$ induced by a spatially varying phase damping channel with a spatially dependent noise parameter, $p(x,y)$, as defined in Eq.\ref{simplechoices}. Although the new map $\vec{S}'(x,y)$ reflects a highly distorted geometry when compared to the initial spherical geometry, the Skyrmion number remains robust at $N\approx0.98$. This is expected since the noise parameter varies smoothly with the spatial coordinates. In Fig.~\ref{fig:PhaseDampSpatial}(b), results for various states with a range of topological numbers, all acted on by the same spatially varying phase damping channel, demonstrate the complete robustness of the Skyrmion topology under the influence of this type of noise, regardless of noise levels or spatial variation.

\section{Global Effects}\label{GlobalF}

In the preceding sections, we have developed a description for noise that affects only local features of the wave function. Under these conditions it is reasonable to expect that noise leaves the topology intact. However, as we will now explain, certain local noise effects can disturb the topology of the wave function. To understand how this can happen, recall that the integer Skyrmion number is the winding number of a map from $S^2$ to $S^2$. One of these spheres represents the space of polarization states, while the other is obtained by compactifying the plane $\mathbb{R}^2$ to a sphere. This compactification is achieved by identifying the points at infinity\footnote{The points at infinity have $\rho = \infty$ and any $\phi$ value.} with the north pole of an $S^2$. Compactification is only possible if the wave function takes the same value for all points at infinity, as it is only in this case that it makes sense to identify this collection of points with a single point, the north pole of a sphere. If the compactification condition is not satisfied, the Skyrmion number need not even be an integer and our scheme for the discretization of entanglement fails.

We will discuss this effect in the simplest possible setting. Our Skyrmion wave functions lead to Stokes parameters of the following generic form
\begin{eqnarray}
S_x&=&\frac{2\sqrt{\rho}\sin(k\phi )}{\rho+1}\cr\cr
S_y&=&\frac{2\sqrt{\rho}\cos(k\phi)}{\rho+1}\cr\cr
S_z&=&\frac{\rho-1}{\rho+1}
\end{eqnarray}
This particular solution has a Skyrmion number of $k$. The important features of these Stokes parameters are the following:
\begin{itemize}
\item[1.] $S_x$ and $S_y$ are responsible for all of the $\phi$ dependence.
\item[2.] The Stokes vector is normalized to 1: $\vec{S}\cdot\vec{S}=1$.
\item[3.] $S_z$ is $-1$ at the origin $\rho=0$ and it tends to $1$ as $\rho\to\infty$.
\end{itemize} 
These general features are all that are needed to discuss the compactification condition and they are all shared with the actual Skyrmion solutions we have considered. After a suitable rotation, it is reasonable to expect that many sensible Skyrmion solutions will have these properties. We can argue that any solution which enjoys these three properties will obey the compactification condition. Indeed, property 3 above tells us that $S_z=1$ for the points at infinity, while property 2 then implies that $S_x=0=S_y$ for these points. Finally, property 1 then establishes that there is no $\phi$ dependence in the Stokes parameters as $\rho\to\infty$ so that we can indeed compactify the plane to S$^2$.

It may sound implausible that noise can affect the behaviour of the wave function at infinity. However, a key insight of \cite{wang2024topological} is that in practice we never work with the entire plane, but rather with some open subset $U \subset \mathbb{R}^2$. The compactification condition then amounts to the requirement that on the boundary of the open subset $U$, the Stokes parameters no longer have any $\phi$ dependence. Noise can disrupt this property by introducing additional $\phi$ dependence. In this case, the plane cannot be compactified to an $S^2$, we lose the interpretation of the Skyrmion wave functions as a map from $S^2$ to $S^2$, and generally, there is no reason to expect that the Skyrmion number computed using (\ref{SkNumb}), will give an integer. 

\begin{figure}[t!]
\includegraphics[width=1.0\linewidth]{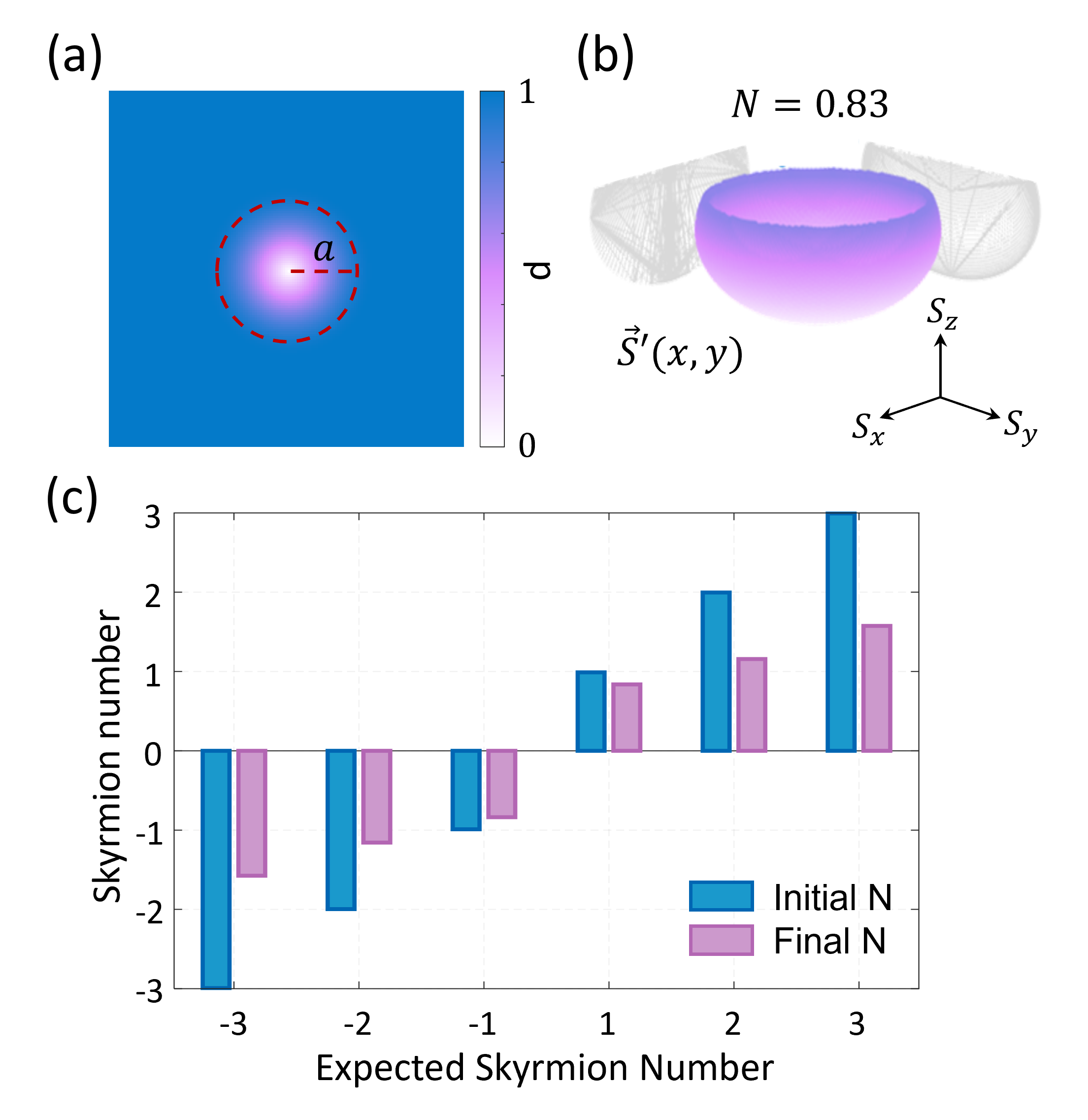}
\caption{{\bf Violation of the Skyrmion compactification condition.} (a) A spatially varying depolarizing noise function where $p<1$ inside the outlined region (dashed red circle) and $p=1$ outside of it. (b) The noisy Skyrmion map, $\vec{S}'(x,y)$, for $N=1$ after normalization, showing partial coverage of the Poincaré sphere, with a calculated Skyrmion number of $N\approx0.83$. (c) Numerically calculated Skyrmion numbers before and after passing through the noisy channel for various topologies, $N \in \{-3, -2, -1, 1, 2, 3\}$, demonstrating the loss of integer discretization in the topological signal.} 
\label{fig:GlobalDep}
\end{figure}

We can also examine a scenario where both conditions \textbf{2} and \textbf{3} are violated in a practical setting, as illustrated in Fig.\ref{fig:GlobalDep}. In this case, depolarizing noise reaches its maximum value, $p=1$, at a finite radius, $r=a$, as shown in Fig.\ref{fig:GlobalDep} (a). This situation can occur when the signal-to-noise ratio at the periphery becomes too low to detect any meaningful signal, making the state appear maximally mixed in which case we can't distinguish the vector of Stokes parameters from the null vector. Within the region where $r<a$, we can maintain $\vec{S}'\cdot\vec{S}'=1$ after normalization. However, outside this region, for $r\geq a$, this condition can no longer be satisfied, as $\vec{S}'\cdot\vec{S}'=0$. This breakdown is also depicted in Fig.\ref{fig:GlobalDep} (b), which shows a clear loss of coverage of the Poincaré sphere by the map $\vec{S}'$. Such violations generally lead to maps with non-integer Skyrmion numbers, as further illustrated in Fig.\ref{fig:GlobalDep} (c), where states with different topologies are passed through the same channel.

%\section{Experimental Results}
%\begin{figure}[h]
%\includegraphics[width=1.0\linewidth]{RoughFigures/RoughExpFigure.PNG}
%\caption{\PO{Placeholder for experiment. The idea here is to not only show the experiment (it should be noted that this is the first experiment with the altered SPC for improved results), but to emphasize the approach to produce noise. We take a statistical ensemble approach where the experiment runs for a long time so as to construct and measure entangled photon pairs in different configurations. The final density matrix is then an average of all of these configurations. This figure should also include some characterization measurements. As a characterization, we can report the construction of each component density matrix with high fidelity.}}
%\label{fig:Exp}
%\end{figure}

%\begin{figure}[h]
%\includegraphics[width=1.0\linewidth]{RoughFigures/RoughResultsStructure.PNG}
%\caption{\PO{Placeholder for results. Results will consist of plot showing skyrmion number and entanglement witnesses against increasing noise parameters. Theory (lines) and experimental points. Some examples could be chosen to show how their density matrices transform, and how the geometry of the Hilbert space has changed.}}
%\label{fig:Results}
%\end{figure}

\section{Conclusions}\label{conclusion}

In this paper, we have examined a scheme that discretizes entanglement by leveraging the non-trivial topology of the entangled wave function. Our primary focus has been to investigate the robustness of the resulting discrete signal against noise.

Discrete signals are inherently robust against noise because the noise must induce a transition between discrete values before any effect is observed. Moreover, since topology is encoded into the global features of the wave function, noise that only induces local changes is unlikely to disturb the topology. Intuitively, transitions between discrete values are not easily effected by noise, allowing our scheme to benefit from a natural topological noise rejection property.

For non-depolarizing sources of noise our description employs a quantum channel with a single Krauss operator. In this scenario, we constructed a rigorous argument based on homotopies between the noisy and noise-free density matrices, which establishes the topological noise rejection property. This theoretical argument was corroborated by explicit numerical analysis.

Quantum channels that describe depolarizing noise inherently involve a sum over Krauss operators rendering a homotopy analysis inapplicable. The basic point is that each term in the Krauss operator sum contributes to the Stokes parameters and since the Skyrmion number (\ref{SkNumb}) is cubic in the Stokes parameters there will be interference effects between the contributions from different channels. Arguing that homotopies exist for the individual channel terms does not constrain their interference so that the existence of homotopies no longer implies the invariance of the Skyrmion number. Nevertheless, numerical analysis can still be performed. Our findings confirm that the Skyrmion number remains robust even under the influence of depolarizing noise.

Finally, the integer Skyrmion number is the winding number of a map from $S^2$ to $S^2$. The base sphere is obtained by compactifying the plane $\mathbb{R}^2$ and the target sphere is the space of polarization states. Compactification of the plane identifies the points at infinity (points with $\rho = \infty$ and any $\phi$) as a single point. This is only possible if the wave function becomes independent of $\phi$ as $\rho\to\infty$. If the wave function fails to meet this condition, the invariant computed using (\ref{SkNumb}) can no longer be considered a Skyrmion number and is not generically integer: the discretization of entanglement through the use of topology fails.

The results of this paper establish the conditions under which the Skyrmion topology is robust to the influence of noise. Our model for the noise is novel and we believe it gives a solid foundation for further investigations into the topological resilience of carefully defined entanglement observables and we hope our findings will motivate further developments in this exciting field.

\section*{Acknowledgements}
This work was supported by the South African National Research Foundation/CSIR Rental Pool Programme. The research of RdMK is supported by a start up research fund of Huzhou University, a Zhejiang Province talent award and by a Changjiang Scholar award. BQL is supported by the National Natural Science Foundation of China under Grant No.~12405058 and by the Zhejiang Provincial Natural Science Foundation of China under Grant No.~LQ23A050002.
%
%%%%%%
\section*{Author contributions}
R.M.K. and B.Q.L. provided the theoretical framework. R.M.K., B.Q.L. and P.O. performed the numerical simulations. All authors contributed to the writing of the manuscript and analysis of data. The idea for the project was conceived by R.M.K., P.O., I.N. and A.F. R.M.K. and A.F. supervised the project.

%%%%%%
\section*{Competing Interests}
The authors declare no competing interests.

\section*{Data availability}
The data are is available from the corresponding author on request.

%%%%%%%%%%%%%%%%%%%%%%%%%%%%%%%%%%%%%%%%%%%%%%%%%%%
%%%% References
%%%%%%%%%%%%%%%%%%%%%%%%%%%%%%%%%%%%%%%%%%%%%%%%%%%

\end{document}